\documentclass[10pt]{article}
\usepackage{amsmath}
\usepackage{graphicx}
\usepackage{amsfonts}
\usepackage{amssymb}
\usepackage{epsf}
\usepackage{latexsym}

\def\mh{\mbox{h}}

\def\d{\mbox{d}}

\def\ml{\mbox{l}}

\def\be{\begin{equation}}
\def\beq{\begin{equation}}
\def\ee{\end{equation}}
\def\eeq{\end{equation}}
\def\bea{\begin{eqnarray}}
\def\eea{\end{eqnarray}}

\def\pa{\partial}
\def\d{\textrm{d}}

\def\hat{\widehat}
\def\sa{\mbox{\scriptsize a}}

\def\sd{\mbox{\scriptsize d}}
\def\sa{\mbox{\scriptsize a}}

\def\sd{\mbox{\scriptsize d}}

\def\sh{\mbox{\scriptsize h}} 
\def\si{\mbox{\scriptsize i}}
 
\def\sk{\mbox{\scriptsize k}}
\def\sll{\mbox{\scriptsize l}}  
\def\sm{\mbox{\scriptsize m}}
\def\sn{\mbox{\scriptsize n}} 
\def\so{\mbox{\scriptsize o}}

\def\sr{\mbox{\scriptsize r}}

\def\st{\mbox{\scriptsize t}}
\def\su{\mbox{\scriptsize u}}

\def\sB{\mbox{\scriptsize B}}

\def\sJ{\mbox{\scriptsize J}}
\def\sK{\mbox{\scriptsize K}}

\def\sW{\mbox{\scriptsize W}}
 
\def\sY{\mbox{\scriptsize Y}} 
 
\def\se{\mbox{\scriptsize e}} 
\def\eph{\mbox{\scriptsize eph}}
 
\def\eph(B){\mbox{\scriptsize emergent(LMB)}}

\def\fE{\mbox{\sffamily E}}
\def\fF{\mbox{\sffamily F}}

\def\fS{\mbox{\sffamily S}}
\def\fT{\mbox{\sffamily T}}
\def\fU{\mbox{\sffamily U}}
\def\fV{\mbox{\sffamily V}}
\def\fW{\mbox{\sffamily W}}

\def\NSI{Na\"{\i}ve Schr\"{o}dinger Interpretation }
\def\CPI{Conditional Probabilities Interpretation }

\begin{document}

\begin{titlepage}

\begin{center}

\vspace{.3in}

{\Large{\bf THE PROBLEM OF TIME IN QUANTUM GRAVITY}} 

\vspace{.3in}

{{\bf Edward Anderson$^*$}}

\vspace{.3in}

\noindent {\sl DAMTP, Cambridge, U.K.}

\noindent  and 

\noindent {\sl Departamento de F\'{\i}sica Te\'{o}rica, Universidad Autonoma de Madrid.}  

\end{center}

\vspace{.3in}


\begin{abstract}

The problem of time in quantum gravity occurs because `time' is taken to have a different meaning in 
each of general relativity and ordinary quantum theory.  
This incompatibility creates serious problems with trying to replace these two branches of physics with 
a single framework in regimes in which neither quantum theory nor general relativity can be neglected, 
such as in black holes or in the very early universe. 
Strategies for resolving the Problem of Time have evolved somewhat since Kucha\v{r} and Isham's 
well-known reviews from the early 90's.  
These come in the following divisions 
I) time before quantization, such as hidden time or matter time. 
II) Time after quantization, such as emergent semiclassical time. 
III) Timeless strategies of Type 1: na\"{\i}ve Schr\"{o}dinger interpretation, conditional 
probabilities interpretation and various forms of records theories, and 
Type 2 `Rovelli': in terms of evolving constants of the motion, complete observables and partial 
observables.  
IV) I argue for histories theories to be a separate class of strategy.    
Additionally, various combinations of these strategies have begun to appear in the literature; 
I discuss a number of such. 
Finally, I comment on loop quantum gravity, supergravity and string/M-theory from the problem of time 
perspective.  

\end{abstract}


\vspace{.3in}
 
\mbox{ }




\noindent{\bf PACS numbers 04.60.Ds,Kz and 04.20.Cv,Fy}

\vspace{2in}

\noindent$^*$ ea212@cam.ac.uk

\end{titlepage}

\section{Introduction to the Problem of Time}\label{Sec: POTIntro}

\subsection{The Problem of Time}

The Problem of Time \cite{Wheeler, K81, K91, K92, I93, K99, Kieferbook, Rovellibook, Smolin08} occurs 
because `time' takes a different meaning in each of ordinary Quantum Theory (QT) and General Relativity 
(GR).  
This incompatibility creates serious problems with trying to replace these two branches of physics with 
a single framework in regimes in which neither QT nor GR can be neglected, as occurs in black holes or 
in the very early universe. 
Resolving this incompatibility is of clear importance if theoretical physics is to form a coherent 
whole. 
Study of the Problem of Time is also important toward acquiring more solid foundations for the 
gradually-developing discipline of Quantum Cosmology (see e.g. \cite{H03}). 
I next consider a number of facets of the Problem of Time (and various technical issues that become 
entwined with these).

\subsection{The Frozen Formalism Facet of the Problem of Time} 

One notable facet of the Problem of Time shows up in attempting canonical quantization of GR (or many 
other gravitational theories that are likewise background-independent).  

\mbox{ } 

\noindent {\bf Frozen Formalism Facet} The canonical approach at the classical level gives a constraint 
that is quadratic in the momenta but contains no linear dependence on the momenta.  
For GR, this is the Hamiltonian constraint,\footnote{Let $\Sigma$ be some fixed spatial topology, taken 
to be a compact without boundary one for simplicity.  
Then $h_{ab}$ is a spatial metric on $\Sigma$, with determinant $h$, conjugate momentum $\pi^{ab}$, 
covariant derivative $D_a$ and Ricci scalar Ric(x; $h$]. 
The many possible $h_{ab}$ on $\Sigma$ form the redundant configuration space Riem($\Sigma$).  
$G^{abcd}  := h^{ac}h^{bd} - h^{ab}h^{cd}$ is the GR configuration space metric (N.B. that this is 
indefinite), with determinant $G$, inverse $G_{abcd} = h_{ac}h_{bd} - h_{ab}h_{cd}/2$ (the undensitized 
form of the DeWitt supermetric) and Ricci scalar Ric($h;G$].  
${\cal H}^{\mbox{\tiny matter}}$ is proportional to the energy density of the universe model's 
matter, and ${\cal H}_i^{\mbox{\tiny matter}}$ is proportional to the momentum flux of the 
universe model's matter. 
$\Lambda$ is the cosmological constant and $\Psi$ is the wavefunction of the universe.  
I use round, square and $(\mbox{ }; \mbox{ }]$ brackets for function, functional and mixed dependence 
(function; functional].
I use fundamental units.}
\beq
{\cal H} := G_{abcd}\pi^{ab}\pi^{cd}/\sqrt{h} - \sqrt{h}\{\mbox{Ric}(x;h] - 2\Lambda\} + 
{\cal H}^{\sm\sa\st\st\se\sr} = 0  \mbox{ } .  
\eeq
Then promoting an equation with a momentum dependence of this kind to the quantum level does not give 
a time-dependent wave equation such as (for some notion of time $t$ and some quantum Hamiltonian 
$\widehat{H}$)
\beq
i\pa\Psi/\pa t = \widehat{H}\Psi
\eeq
as one might expect, but rather a stationary, i.e. frozen, i.e. timeless equation 
\beq
\widehat{\cal H}\Psi = 0 \mbox{ } .  
\label{calH}
\eeq
In the case of GR, this is the {\it Wheeler--DeWitt equation} (WDE); in more detail,
\beq
\hat{\cal H}\Psi := - ``\frac{1}{\sqrt{{G}}}\frac{\delta}{\delta h^{{ab}}}
\left\{
\sqrt{{G}}{G}^{abcd}\frac{\delta\Psi}{\delta h^{{cd}}}
\right\} 
- \xi \,\mbox{Ric}(h;G]\mbox{"}\,\Psi -  \sqrt{h}\mbox{Ric}(x;h]\Psi + {\sqrt{h}2\Lambda   }\Psi  
+ \hat{\cal H}^{\mbox{\scriptsize matter}}\Psi = 0  \label{WDE} \mbox{ } .     
\eeq
This suggests, in apparent contradiction with everyday experience, that noting at all {\sl happens} in 
the universe! 
Thus one is faced with having to explain the origin of the notions of time in the laws of physics that 
appear to apply in the universe; this paper reviews a number of strategies for such explanations.   
[Moreover timeless equations such as the WDE apply {\sl to the universe 
as a whole}, whereas the more ordinary laws of physics apply to small subsystems {\sl within} the 
universe, which does suggest that this is an apparent, rather than actual, paradox.]

N.B. that, whilst considering special relativity changes {\sl the form} of the 
time-dependent wave equations involved, {\sl the passage from Newtonian theory to special relativity 
merely swaps Newton's notion of absolute time for another absolute notion -- that associated with the 
timelike Killing vector of Minkowski spacetime}.  
In this sense, it is the passage to {\sl general} relativity that is the one which brings about a 
Problem of Time (see \cite{Kiefer09bis} for more).
This is closely tied to how GR has two aspects: it is not only a theory of gravitation, but also, 
following Einstein, a way of freeing physics from absolute structures. 
This second aspect, moreover, is quite clearly not specific just to the field equations of GR; a wider 
range of theories with different field equations also possess such a `background-independent' or 
`diffeomorphism invariant' aspect.

One follow-on from the Frozen Formalism Facet as manifested by a frozen equation like the 
WDE of GR is the Hilbert Space/Inner Product Problem, i.e. how one is to turn the 
space of solutions of that frozen equation into a Hilbert space.

\subsection{Some accompanying complications in attempting canonical quantization of gravity}

1) The reader will have noticed unexplained inverted commas in formula (\ref{WDE}).  
These indicate that the Wheeler-DeWitt equation has, in addition to the Problem of Time, various 
technical problems, including operator ordering issues, regularization problems, and whether functional 
differential equations are meaningful.   

\mbox{ }

\noindent 2) The preceding SSec's page's `plain quantum geometrodynamics' scheme corresponds to 
considering the canonical 

\noindent
commutation relation $[\widehat{h}_{ij}(x), \widehat{\pi}^{ab}(x^{\prime})] = 
i\delta_i^a\delta_j^b\delta^3(x,x^{\prime})$.  
However, there is also an alternative `affine quantum geometrodynamics' scheme \cite{Klauder, Affine} 
in which, so as to keep on imposing the classical spacelike metric condition det($h$) $>$ 0 at the 
quantum level, one considers, rather, the affine commutation relation 
$$
[h_{ij}(x),\pi_{l}
\mbox{}^m(x^{\prime})] = i h_{ij}\delta_m\mbox{}_j(x)\delta^3(x, x^{\prime}) \mbox{ } .
$$  
[Strictly, the right hand sides in this paragraph are symmetrized over $i$, $j$.]
Others may prefer to pass from geometrodynamical variables to Ashtekar Variables, or to consider the 
supesymmetric version of GR (Supergravity) or some form of String Theory.
The Problem of Time persists throughout all these theories (at a deep enough level), but how various 
strategies to deal with it fare differs from theory to theory. 
This is the subject of much of the Conclusion.  

\mbox{ } 

\noindent 3) The canonical approach to GR also gives the {\it momentum constraint}, 
\beq
{\cal H}_i   := - 2D_{j}{\pi^{j}}_i  + {\cal H}_i^{\sm\sa\st\st\se\sr} = 0 \mbox{ } .
\eeq
Unlike the quadratic constraint, the momentum constraint is purely linear in the momenta. 
Furthermore, the momentum constraint embodies invariance under spatial 3-diffeomorphisms Diff($\Sigma$).  
Taking this into account, one passes from Riem($\Sigma$) to the less redundant/more reduced 
configuration space Superspace($\Sigma$) = Riem($\Sigma$)/Diff($\Sigma$) of spatial 3-geometries.
Promoted to the quantum level, the momentum constraint takes the form 
\beq
\widehat{\cal H}_i \Psi  := - 2h_{ik}D_{j}\frac{\delta}{\delta h_{jk}} \Psi + 
\widehat{\cal H}_i^{\sm\sa\st\st\se\sr}\Psi = 0 \mbox{ } .
\label{calHi}
\eeq

\noindent 4) The canonical approach considers a sequence of spatial 3-surfaces that constitute a 
foliation of spacetime.  
Moreover, invariance under 4- (rather than 3-)diffeomorphisms Diff(${M}$) of the spacetime $M$ is 
altogether harder to find/account for in canonical approaches to GR. 
In particular, the `missing' diffeomorphisms are not directly accountable for via the presence of the 
Hamiltonian constraint.  
This is reflected e.g. in the Dirac algebra of the Hamiltonian and momentum constraints being distinct 
from, and much less mathematically tractable than, the algebra of spacetime 4-diffeomorphisms. 
In particular, this algebra involves not structure constants but more general structure functions that 
depend on the 3-metric variable.  
At least this algebra has the useful property of closing at the classical level without further 
constraints becoming necessary to do so.
Also, in classical GR, one can foliate spacetime in many ways, each corresponding to a different choice 
of timefunction.  
Thus time in classical GR is `many-fingered', with each finger `pointing orthogonally to each possible 
foliation.  
And, furthermore, classical GR has the remarkable property of being foliation independent, so that 
going between two given spatial geometries by means of different foliations in between produces the 
same region of spacetime and so the same answers to whatever physical questions can be posed therein.

\mbox{ }

\noindent Finally, note that the momentum constraint often becomes entwined in technical problems 
becoming Problem of Time strategies; many such entwinings furthermore specifically involve 
diffeomorphisms (see Sec 1.5).

\subsection{Further Facets of the Problem of Time}\label{SSSec: Facets}

Over the past decade, it has become more common to argue or imply that the Problem of Time {\sl is} the 
Frozen Formalism Problem.  
However, a more long-standing point of view \cite{K92, I93} that the Problem of Time contains a number 
of further facets; I argue in favour of this in this paper.  
It should first be stated that the problems encountered in trying to quantize gravity largely interfere 
with each other rather than standing as independent obstacles.
Kucha\v{r} talked of this in \cite{K93} as a `many gates' problem, in which one attempting to enter the 
gates in sequence finds that they are no longer inside some of the gates they had previously entered. 
(The object being described is presumably some kind of enchanted castle, or, at least, a topologically 
nontrivial one).  
Additionally, the various of these problems that are deemed to be facets of the Problem of Time do bear 
conceptual and technical relations that makes it likely to be advantageous to treat them as parts of a 
coherent package rather than disassembling them into a mere list of problems to be addressed piecemeal.
For, these facets arise from a joint cause, i.e. the mismatch of the notions of time in GR and QT.

As such, I suggest the following further parable the Problem of Time.  
One notices that the physical equations get frozen.  
One reaction is to set about trying to unfreeze them.
However, another perspective is that an Ice Dragon \cite{GRRM} may be on the loose, a beast not that not only 
possesses the `freezing breath' of the Frozen Formalism but which is the joint cause of other observed 
devastations, through coming simultaneously equipped with 1) `many-fingered claws' on four `limbs' 
that are the somewhat interlinked 
\noindent i) Functional Evolution Problem,  
\noindent ii) Foliation Dependence Problem, 
\noindent iii) Multiple Choice Problem, and 
\noindent iv) Sandwich Problem, 
\noindent 2) the `thunderbolt tail' of the Global Problem of Time,   
\noindent 3) the `scaled armour' of the Spacetime Reconstruction Problem, and   
\noindent 4) the `wings' of the Problem of Observables.  
\noindent Nor do the attributes of the Ice Dragon take an entirely fixed form.  
E.g. in path-integral rather than canonical approaches, one has to reckon with a Measure Problem 
instead of some of the difficulties of interpreting a frozen wave equation.   
Now, setting about unfreezing the physics is likely more straightforward than defeating an Ice Dragon.
Some might even claim to be defeating the Ice Dragon by just unfreezing the physics, by taking its 
rather feared `Problem of Time' name and rebranding this name to mean just its `Frozen Formalism 
Facet.'
However, this amounts to leaving oneself open to the `claws', `tail', `scales' and `wings' by choosing 
not to take any advantage of the presence of a joint cause -- the mismatch of the notion of time in GR 
and QT -- as a warning that these, too, would also be expected to be present as obstacles needing 
overcoming before one's quantum gravity program has genuine long-term viability. 
(Though it may take long for this to be demonstrated, through requiring detailed analysis of difficult 
and diverse quantum-gravitational calculations). 
I next outline the nature of these other facets of the Problem of Time.

\subsection{Outline of the other Facets}\label{SSec: Facets}

The {\bf Sandwich Problem} is already present at the classical level and gets involved in some of the 
approaches that start with manipulations at the classical level prior to quantizing.
It consists in the following. 
The generating function for the canonical transformation corresponding to the evolution of the 
gravitational field needs to be a function of the initial and final slices' metrics in the classical 
configuration representation.    
However, this is not a unique procedure in the absense of a resolution of the Thin Sandwich Conjecture 
\cite{TSC}. 
 
\mbox{ } 

\noindent
For all that one would like a quantization of GR to retain the nice classical properties of 
refoliation invariance and closure of the constraint algebra, there is no longer an established way 
of guaranteeing these properties at the quantum level.  
Thus there is a {\bf Foliation Dependence Problem}.  
There is also a {\bf Functional Evolution Problem}: the quantum-level problem that the commutator of 
constraints is capable.\footnote{Dirac \cite{Dirac} 
states that one requires luck to avoid this sort of problem, which is also very widely known as an 
{\sl anomaly}.}  
%
In the present case of the quantum counterpart of the Dirac algebra of GR constraints, this is a 
time issue due to the meaning of these constraints.
In particular, non-closure here is a way in which the Foliation Dependence Problem can manifest itself, 
through the non-closure becoming entwined with details of the foliation. 
[Non-closure is also entwined with the Operator Ordering Problem, since changing the ordering gives 
additional right-hand-side pieces not present in the classical Poisson brackets algebra.]  

\mbox{ }

\noindent
The {\bf Multiple Choice Problem} is the purely quantum-mechanical problem that different choices of time variable may give 
inequivalent quantum theories.  
It is a subcase of how making different choices of sets of variables to quantize may give inequivalent 
quantum theories, which follows from the Groenewold--van Hove theorem \cite{GVH}.  
\noindent Foliation Dependence is one of the ways in which the Multiple Choice Problem can manifest 
itself.
Moreover, the Multiple Choice Problem is known to occur even in some finite toy models. 
It has been suggested that one way out of the Multiple Choice Problem is to specify the lapse $N$ and 
shift $N_i$ a priori (e.g. in \cite{unim-SUGRA}). 
However, such amounts to foliation fixing, which is not what done in practise, nor desirable in 
principle, since one would like to be able to ascribe physical meaning to whichever foliation (and 
consider e.g. whether two distinct foliations between two given hypersurfaces match in their physical 
predictions at the quantum level).  
Also, foliation issues are not the only possible source of the Multiple Choice Problem.  

\mbox{ }

\noindent The {\bf Global Problem of Time} is already present at the classical level. 
It consists of the separation into true and embedding (space frame and timefunction) variables' having 
a capacity for being globally impossible, for reasons in close parallel to the Gribov ambiguity in 
Yang--Mills theory.  
The Global Problem of Time moreover subdivides into a part by which such a split cannot be defined on a 
given hypersurface, and a part by which such a split can be defined but cannot be indefinitely 
continued as one progresses along a foliation.

\mbox{ } 

\noindent The {\bf Spacetime (replacement or reconstruction) Problem.}  
Internal space or time coordinates to be used in the conventional classical spacetime context need to be 
scalar field functions on the 4-manifold $M$.    
In particular, these do not have any foliation dependence.
However, the canonical approach to GR uses functionals of the canonical variables, and which there is 
no a priori reason for such to be scalar fields of this type.  
Thus one is faced with either finding functionals with this property, or coming up with a new means 
of reducing to the standard spacetime meaning at the classical level.   
There are further issues involving properties of spacetime being problematical at the quantum level.  
QT implies fluctuations are unavoidable, but now that this amounts to fluctuations of geometry, these 
are moreover too numerous to fit within a single spacetime (see e.g. 
%
%
\cite{Wheeler}).  
Thus (something like) the superspace picture (considering the set of possible 3-geometries) might be 
expected to take over from the spacetime picture at the quantum level.  
It is then not clear what becomes of causality (or of locality, if one believes that the quantum 
replacement for spacetime is `foamy').  

\mbox{ } 

\noindent The {\bf Problem of Observables}.  This involves construction of a sufficient set of 
observables for the physics of one's model, which are then involved in the model's notion of 
evolution.  

\mbox{ } 

\noindent 
One further issue that is usually considered to be not part of the Problem of Time mismatch but rather 
a further time problem is the {\bf Arrow of Time}.  
This is a further issue since it concerns not ``what happens to the notion of time upon jointly considering 
QT and GR" but rather ``why is there a (theory-independent and in practise observed) consistent distinction 
between past and future?" 
Quantum Gravity or Quantum Cosmology are then sometimes evoked in attempts to better explain this 
than one has managed to do with more everyday physical theories.

\subsection{Strategies for the Problem of Time}

Over the years, many conceptual strategies have been put forward to resolve the Problem of Time.  
However, none of those proposed to up to 1993 worked upon detailed examination \cite{K92, I93}, while 
not all work since has been investigated in such detail.
Some of the differences in Problem of Time strategies result from what are (see e.g. \cite{Denbigh}) two 
well-established and conflicting philosophical positions concerning time  

\noindent A) that time is fundamental, or 

\noindent B) that time should be eliminated from one's conceptualization of the world.  

\noindent
These two perspectives can be related in various ways to schemes for the world, and the scientific 
enterprise therein, in which question-types involving `becoming' fundamentally make sense, or only 
those involving `being'. 

\noindent In Kucha\v{r} \cite{K92} and Isham \cite{I93}, classification of the Problem of Time 
strategies, another element is whether the identification of what is the time is to be done before or 
after quantization (if time is to be identified at all).  
Thus strategies for the Problem of Time can be divided up as follows.  

\mbox{ } 

\noindent {\bf Tempus Ante Quantum} schemes (Sec 2) follow A) by, prior to quantization, having a 
fundamental time in their general case.  
Such strategies include hidden time in pure geometrodynamics, a matter time arising from the adjunction 
of matter to geometrodynamics, and unimodular gravity (in which a dynamical cosmological constant 
provides a time-standard).  

\mbox{ } 

\noindent {\bf Tempus Post Quantum} schemes (Sec 3) follow B) insofar as there is no time in 
general at the classical level, but follow A) insofar as one restricts attention to cases for 
which a suitable notion of time does then emerge at the quantum level.
Such strategies include superspace time and an associated Klein--Gordon-type inner product, Third 
Quantization and the Emergent Semiclassical Time Approach.  

\mbox{ } 

\noindent {\bf Tempus Nihil Est} schemes, follow B) insofar as they deny a role for time 
and then see how far one can do physics without.  
Such strategies included, by the early 90's, the Na\"{\i}ve Schr\"{o}dinger Interpretation, the 
Conditional Probabilities Interpretation (both among Sec 4's `Type 1  
Tempus Nihil Est Approaches'), Histories Theory (Sec 5), and a rather distinct 
Type 2 `Rovelli' Approach involving evolving constants of the motion (Sec 7).

\mbox{ } 

\noindent I note moreover that Histories Theory supplants the notion of time by the notion of history. 
This has somewhat of a different status, insofar as i) one passes from configuration space and the 
associated phase space being regarded as primary and being subjected to quantization to it being 
histories that are regarded as primary and subjected to quantization. 
ii) Histories Theory is a spacetime-first approach, whilst all the other approaches that Isham and 
Kucha\v{r} term `Tempus Nihil Est' are space-first approaches.
Due to this distinction, I myself consider Histories Theory to lie outside of my own use of `Tempus 
Nihil Est', as a fourth division dubbed {\bf Non Tempus Sed Historia}.

\mbox{ } 

\noindent
(Spacetime primality versus spatial primality is also an issue in Tempus Ante Quantum and Tempus Post 
Quantum strategies; spacetime primality is linked to the wish to maintain, or recover, spacetime 
general covariance/spacetime diffeomorphism invariant formulations.  
Some approaches also assume strong preference for the primality of one of spacetime or space.  
Bias towards spacetime approaches may well come from the spacetime perspective is a very fruitful 
approach to classical GR and in quantum particle physics in a Minkowskian background.  
However, it is far from clear if such successes will extend to the realm of Quantum Gravity 
\cite{Wheeler}.  
This `spacetime versus space' point will be expanded on in various Secs below.)

\mbox{ }

The main developments since the early 90's, which are covered mostly in the second half of this 
paper, are as follows. 

\noindent 1) Histories Theory has diversified (Sec 5).  

\noindent
2) Each side of Tempus Nihil Est's split into Type 1 and Type 2 `Rovelli' has further expanded and 
diversified.   
For Type 1, various forms of timeless Records Theory have appeared (Sec 4), including Barbour's, Page's 
and mine, and also Records Theory within Histories Theory (work of Halliwell \cite{H99} 
building on work of Gell-Mann and Hartle \cite{GMH}, see Secs 5 and 6). 

\noindent
3) For Type 2, most of the advances in it (by Rovelli, Thiemann and Dittrich, see Sec 7 and 8) are tied 
to further developments in Loop Quantum Gravity (though many of the other types of Problem of Time 
strategy are also considered in this arena, see Sec 8.3.) 

\noindent 4)
Additionally, one can consider further combinations of some of the types of strategy; 
see e.g. Secs 6 and 8. 

\noindent 
[There have also been some advances \cite{BK97,HKG06,New6} in the matter time approach to Tempus 
Ante Quantum.] 

\noindent 
Thus, all in all, one could view the present situation as the 22 strategies in Figure 1.    

\mbox{ } 

Bear in mind throughout that there are issues throughout these approaches as regards the extent to 
which each candidate time involved (actual or emergent) resembles the time of QT and that of GR.  
The issue of good and bad timefunctions/clock variables/clocks is quite an important theme in the 
present paper.  
E.g. \cite{UW89, Hartle96, SemiclI, Zeh09} have a number of criteria for, and examples of, good and bad 
timefunctions. 
 
\mbox{ }  

Also, certainly each of the 10 strategies covered by Isham and Kucha\v{r} (Fig 1) has problems if 
examined in detail \cite{K92, I93}. 
Taking these as precedent, I consider it likely that subsequent strategies will also have great 
difficulty in being genuinely satisfactory; this however remains undemonstrated in some cases.  
A resolution for the Problem of Time being acceptable would entail it to work in the general case for 
a fully viable theory of gravitation rather than just working for special cases or toy models.  
I in no way claim that this paper is exhaustive as regards what problems various strategies have 
(e.g. \cite{K92, I93, Hartle, K99, forth} cover plenty more such issues).    

\mbox{ }

Nor is the present article's coverage exhaustive -- I expect to provide more thorough reviews (and 
further original investigations of some of the strategies) over the next 10 years.  
In particular, my accounts in the Conclusion of the Problem of Time in the Ashtekar Variables, 
Supergravity and String/M-theory settings are not exhaustive. 
Consult the Appendix if unfamiliar with any toy models mentioned in the main text.

{            \begin{figure}[ht]
\centering
\includegraphics[width=1.0\textwidth]{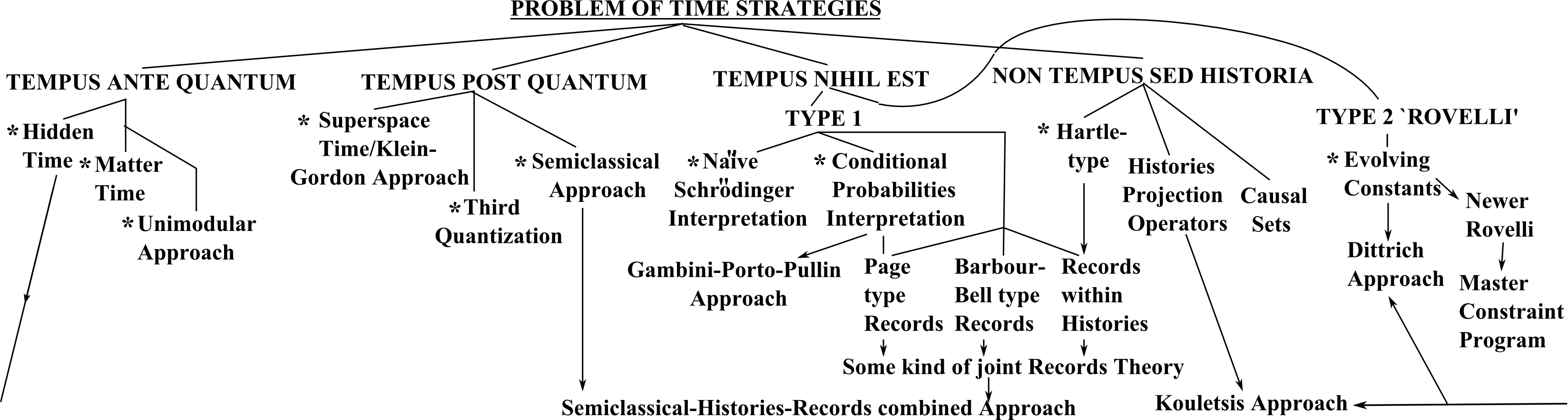}
\caption[Text der im Bilderverzeichnis auftaucht]{        \footnotesize{The various branches of 
strategy and their relations, in a diagram with `cylindrical topology'. 
* indicates the 10 strategies covered by Kucha\v{r} and Isham's reviews \cite{K92, K93}.  
This paper's order of presentation is roughly from left to right.}         }
\label{Fig1}\end{figure}          }

\section{Tempus Ante Quantum Strategies}

{\bf Ante Postulate}.  There is a fundamental time to be found at the classical level for the full 
(i.e. untruncated) classical gravitational theory (possibly coupled to suitable matter fields).

\subsection{The simple but blocked-off Jacobi--Barbour--Bertotti implementation} 

Barbour's relational approach to physics gives rise to a Jacobi--Barbour--Bertotti (JBB) time 
$t^{\sJ\sB\sB}$ such that $\pa/\pa t^{\sJ\sB\sB} = \sqrt{\fU/\fT}\pa/\pa\lambda$. 
$\fT$ is the kinetic term of the theory (assumed to be purely quadratic) and $\lambda$ is a meaningless 
label time, cancelling with the other $\lambda$'s present in $\fT$.   
In the context of relational particle mechanics (RPM's, see Appendix A.7), $\fU = \fE - \fV$, for $\fE$ 
the total energy and $\fV$ the potential; it arises as a simplifier of the relational approach's 
classical equations and amounts to a recovery of Newtonian time from relational first 
principles.  
In the context of GR, $\fU = \mbox{Ric}(x;h] - 2\Lambda - {\cal H}^{\sm\sa\st\st\se\sr}$, and 
$\sqrt{\fU/\fT} = 1/N$, for $N$ the lapse arising in an emergent sense; $t^{\sJ\sB\sB}$ now amounts to, 
in various contexts, recovery of proper time and cosmic time.  
Unfortunately, while this approach yields a time function, it does not in any way unfreeze the GR 
Hamiltonian constraint or its energy constraint RPM analogue.  
However, this makes contact with another strategy (the Semiclassical Aproach, Sec 3.3) in which this 
same timestandard emerges in an an unfreezing context.

\subsection{The also-simple but likewise blocked-off superspace time implementation}\label{SSSec: HSTA}

Take the indefinite direction in superspace to be picking out a timefunction, in parallel to how 
the indefinite direction in Minkowski space does for e.g. Klein--Gordon theory.    
This amounts to one of the momenta on which ${\cal H}$ depends quadratically to be taken to be 
conjugate to a timefunction, leading to a Klein--Gordon type wave equation depending on the double 
derivative with respect to this timefuction.  
This approach is usually billed as Tempus Post Quantum, through choosing to make this identification of 
a time after the quantization, but the approach is essentially unchanged if this identification is made 
priorly at the classical level. 
Regardless of when this identification is made, however, this approach fails, for reasons I defer 
to the discussion in its more habitual Tempus Post Quantum setting (Sec 3.1).  
%

\subsection{Implementation by obtaining a Part-Linear Form}

While the above straightforward schemes fail, it may still be that a classical time exists but happens 
to be harder to find.
We now consider starting one's scheme off by finding a way of solving ${\cal H}$ to obtain a 
part-linear form,  
\beq
p_{T^{\sa\sn\st\se}} + 
H^{\st\sr\su\se}(x; T^{\sa\sn\st\se}(x), q_{\so\st\sh\se\sr}^{\Gamma}(x), 
p^{\so\st\sh\se\sr}_{\Gamma}(x)] = 0 
\mbox{ } ,
\label{PLIN}
\eeq
where $p_{T^{\sa\sn\st\se}}$ is the momentum conjugate to a candidate classical time variable, 
$T^{\sa\sn\st\se}$ which is to play a role parallel to that of external classical time.
By passing as soon as possible to having an object that plays such a role, Tempus Ante Quantum can be 
viewed as the most conservative family of strategies \cite{I93, T97}.
$H^{\st\sr\su\se}$ is then the `true Hamiltonian' for the system.  
Given such a parabolic form for ${\cal H}$, it becomes possible to apply a conceptually-standard 
quantization that yields the time-dependent Schr\"{o}dinger equation 
\beq
i\pa\Psi/\pa T^{\sa\sn\st\se} = \hat{H}_{\st\sr\su\se}(x; T^{\sa\sn\st\se}(x), 
q_{\so\st\sh\se\sr}^{\Gamma}(x), p^{\so\st\sh\se\sr}_{\Gamma}(x)]\Psi \mbox{ } , 
\label{TEEDEE}
\eeq 
with obvious associated Schr\"{o}dinger inner product.
The above scheme either amounts to the accompanying momentum constraint having been reduced away or 
being absent through consideration of a simpler toy model, or requires enlarging the scheme to involve 
solving the 4-vector ${\cal H}_{\mu} := [{\cal H}, {\cal H}_i]$ of constraints to obtain a part-linear 
form 
\beq
P_{{{\cal X}}^{\mu}} + 
H^{\st\sr\su\se}_{\mu}(x; q_{\so\st\sh\se\sr}^{\Gamma}(x), p^{\so\st\sh\se\sr}_{\Gamma}(x), 
{\cal X}^{\mu}(x)] = 0 \mbox{ } .
\label{PLIN2}
\eeq
Here, $P_{{\cal X}^{\mu}}$ are the momenta conjugate to 4 candidate embedding variables 
${\cal X}^{\mu}$, which form the 4-vector $[T^{\st\sr\su\se}, X^i]$, where 

\noindent $X^i$ are 3 spatial frame 
variables. 
Also, $H^{\st\sr\su\se}_{\mu} = [H^{\st\sr\su\se}, \Pi^{\st\sr\su\se}_i]$, where $\Pi_i^{\st\sr\su\se}$ 
is the true momentum flux.
Then one passes to a 4-vector enlargement of the time-dependent Schr\"{o}dinger equation, 
\beq
i \delta\Psi/\delta {\cal X}^{\mu} = \hat{H}^{\st\sr\su\se}_{\mu}(x; q_{\so\st\sh\se\sr}^{\Gamma}(x), 
p^{\so\st\sh\se\sr}_{\Gamma}(x), {\cal X}^{\mu}(x)]\Psi \mbox{ } .  
\label{TEEDEE2}
\eeq 
This implementation, while less straightforward than the preceding SSecs', contains some unblocked cases.
Part-linear forms (7) and (9) occurs for parametrized nonrelativistic particle and parametrized field 
theory toy models respectively \cite{K92}.

\subsection{The hidden time Internal Schr\"odinger Approach}\label{SSec: ISI}

A first such suggestion is that there may be a {\it hidden} alias {\it internal} time \cite{York72, K81, 
K92, I93} within one's gravitational theory itself.  
It is to be found by applying some canonical transformation.
In scheme (7), this sends GR's spatial 3-geometry configurations to 1 hidden time, so 
$T^{\sa\sn\st\se} = T^{\sh\si\sd\sd\se\sn}$ plus 2 `true gravitational degrees of freedom' [which are 
the form the `other variables' take, and are here `physical' alias `non-gauge'].
In scheme (9), the canonical transformation is of the form 
\beq
h_{ij}(x), \pi^{ij}(x)  \longrightarrow Q^{\mu}(x), P_{\mu}(x) ; 
q_{\mbox{\scriptsize true grav dof}}^{\Gamma}(x), p^{\mbox{\scriptsize true grav dof}}_{\Gamma}(x) 
\mbox{ } . 
\label{canex} 
\eeq
Thus one gets a hidden time-dependent Schr\"{o}dinger equation of the form (\ref{TEEDEE}) or 
(\ref{TEEDEE2}) with the above significances attached to the variables.
Within geometrodynamics, some candidates that have been considered are as follows.

\noindent 1) {\it Intrinsic time candidates}, though these have largely lacked in successful development 
\cite{K92}). 
One family of examples involve some scale variable as a time: various parametrizations of scale are the 
scalefactor $a$, $\sqrt{h}$ and $\phi$ such that $a = \phi^2 = \sqrt{h}\mbox{}^{1/3}$, as well as the 
{\it Misner time} $\Omega = -\mbox{ln}\,a$.  
Scale has the advatage of being sharply picked out as different from all other degrees of freedom 
(which are scalefree `shape' degrees of freedom), but, unfortunately, it is not monotonic in 
recollapsing universes.  
 
\noindent 2) {\it Extrinsic time candidates}. 
One among these is probably our best candidate for an internal time in GR: {\it York time} 
\cite{York72, K81, K92, I93}, so I present this example in more detail.   
Perform a size-shape split of GR and then exchange the scale variable $\sqrt{h}$ and its conjugate 
dilational momentum object via a canonical transformation so that the latter is the timestandard,  
\beq
T^{\sY\so\sr\sk} := 2h_{\mu\nu}\pi^{\mu\nu}/3\sqrt{h} 
\mbox{ } \mbox{ } \mbox{ } \mbox{ (proportional to the constant mean curvature)} \mbox{ } .   
\eeq
Then the Hamiltonian constraint is replaced by $p_{\st\sr\su\se} = H^{\st\sr\su\se} = \sqrt{h} = 
\phi^6$. 
Here, $\phi$ is the solution of the conformally-transformed Hamiltonian constraint, i.e. the 
Lichnerowicz--York equation, 
\beq
8\nabla^2\phi = \mbox{Ric}(h)\phi - {\pi_{\alpha\beta}\pi^{\alpha\beta}}/{\sqrt{h}}\phi^{7} + 
{\pi^2}\phi^5/{6\sqrt{h}}   
\label{LYE} \mbox{ } . 
\eeq
(For brevity, I present the vacuum case of this, whilst schematically ignoring the momentum constraint; 
more generally, one would formally solve the whole GR initial-value problem.)    
Then quantization gives 
\beq
i \delta\Psi/\delta t^{\sY\so\sr\sk}  = \widehat{    H^{\st\sr\su\se}    }\Psi 
\mbox{ } .  
\eeq
The obstruction to this particular resolution is that how to solve the complicated quasilinear elliptic 
equation (\ref{LYE}) is not in practice generally known. 
Thus the functional dependence of $H^{\st\sr\su\se}$ on the other variables is not known (or 
at best known implicitly via the solution of such a partial differential equation). 
And thus the quantum `true Hamiltonian' $\widehat{H^{\st\sr\su\se}    }$ cannot be 
explicitly defined as an operator (or, at best, is very prone to having the horrendous 
operator-ordering and well-definedness issues that only knowing its form implicitly entails.)
It is also the case that simpler models than full geometrodynamics give severe difficulties just beyond 
where geometrodynamics has this impasse.    

\noindent Another extrinsic time is {\it Einstein--Rosen time} in the context of cylindrical wave 
midisuperspace models \cite{midicyl}.  


\noindent


\noindent Some problems with this SSec's hidden times are as follows.  

\mbox{ }

\noindent Problem 2.5-1. The canonical transformation (11) is hard to perform in practise. 

\noindent Problem 2.5-2. Having not one but a list of candidates, the Multiple Choice Problem may apply.  

\noindent Problem 2.5-3. The Global Problem of Time is present, as follows.
Torre showed that the right-hand-side space of (\ref{canex}) is a manifold while the left-hand-side 
space is not (due to the occasional existence of Killing vectors). 
This limits the sense in which such a map could hold (it certainly does not hold globally). 
H\'{a}j\'{\i}\v{c}ek and Kijowski \cite{HK99} have furthermore established that such a map is not 
unique.  
As regards globality in time, for GR, existence of constant mean curvature slices and propagation and 
monotonicity for these does hold for some, but not all, examples.    

\noindent Problem 2.5-4. Having looked at a number of toy model examples, it is questionable whether the 
above conceptually-standard quantization procedure is can be done in practise. 
For, even in the absense of the above constructibility impasse for York time (for toy models such as 
RPM's, strong gravity (Appendix A.2) and some minisuperspace models, for which the Lichnerowicz--York 
equation is replaced by a solvable algebraic equation) at least some cases would appear to have 
\cite{Kieferbook, SemiclI} well-definedness and positive-probability issues, all of which is tangled 
up with operator-ordering ambiguities too. 

\noindent Also, two Spacetime Reconstruction issues with this approach are as follows \cite{Kieferbook}.  

\noindent Problem 2.5-5.  
The bubble time version [along the lines of (10)] describes a hypersurface in spacetime only after the 
classical equations have been solved. 
Thus its significance at the quantum level is not understood. 

\noindent Problem 2.5-6.  
$T^{\st\sr\su\se}$ is required to be a spacetime scalar and thus obey $\{ T^{\st\sr\su\se}, {\cal H}\} 
= 0$.
That would look to overrule internal times based to purely geometrodynamical phase space variables.

\subsection{Matter clocks and reference fluids} \label{SSSec: Matt}

There are also non-geometrodynamical internal time candidates: {\it Matter Time Approaches}.  
Here, we attain the part-linear form (\ref{PLIN2}) with $T^{\sa\sn\st\se} = T^{\sm\sa\st\st\se\sr}$ and 
$h_{ij}$ in the role of $q^{\Gamma}_{\so\st\sh\se\sr}$ by extending the set of 
variables from the geometrodynamical ones to include also matter variables coupled to these, which then 
serve to label spacetime events.  
(This matter is perhaps ficticious or idealized, but just possibly physical.)
Then one passes to the corresponding form (\ref{TEEDEE2}). 
Examples are Gaussian reference fluid \cite{Gau, K92} and the reference fluid corresponding to the 
harmonic gauge \cite{New2}.

One can also regard an undetermined cosmological constant $\Lambda$ as a type of reference fluid 
\cite{UW89}.
This is the so-called {\it unimodular gravity} program, which Isham and Kucha\v{r} consider as a separate 
entry among their 10 strategies.  
\cite{NewUni} is a more recent review of this, while \cite{UniGryb} discusses how an approach of this 
type arises from Barbour-type relational considerations.  
This approach gives rise to (\ref{PLIN}) with $\Lambda$ itself in the role of linear momentum; 
this in iself is a problem, as one would not expect a single such variable to cope with the 
multitude of slices.

Some overall advantages of this SSec's strategies, the current SSec gets round Problem 2.5-1 by no 
longer requiring some complex canonical transformation of the geometrodynamical variables themselves; 
Problem 2.5-6 and Torre's part of Global Problem 2.5-3 are also absent.      
However, there are the following problems. 

\mbox{ } 

\noindent Problem 2.6-1. 
A Spacetime Reconstruction Problem remains as regards the hypersurfaces staying spacelike 
\cite{K93}.

\noindent Problem 2.6-2. The first paragraph's strategies suffer from having to make a choice between 
having a time variable and having a reference fluid that obeys the energy conditions.  
Though it remains unclear whether this is a coincidence or a theorem that holds for all (or some wide 
class of) matter fields/gauges (an interesting question).  

\noindent Problem 2.6-3. 
Additionally, the Problem of Observables remains an unsettled issue for this Sec's various approaches 
\cite{K93, K94}.

\mbox{ }

\noindent 
A further type of matter time approach involves additionally forming a quadratic combination of 
constraints for dust \cite{BK94}, more general perfect fluids \cite{BroMa}, and massless scalar fields 
\cite{KR95}.  
The point of such (see also \cite{Markopoulou}) quadratic combinations is that they result in strongly 
vanishing Poisson brackets. 

\mbox{ } 

\noindent Problem 2.6-4. 
Unfortunately, such quadratic combinations do not generate the evolution of the geometric data 
\cite{BK97}.

\noindent 
One can also obtain such a quadratic combination and without Problem 2.6-4 in the case of null dust 
\cite{BK97}.

\noindent Problem 2.6-5. However now form (\ref{PLIN2}) is now no longer attained, so time 
``remains lost" in such a formulation.  

\noindent 
Moreover, \cite{HKG06, New6} do then find an internal time in the spherically symmetric case 
(Appendix A.3) with an ingoing and an outgoing null dust shell, using that this model has an 
equivalence with an anisotropic fluid model.

\noindent Possible Problem 2.6-6.  
Finally, I comment that freedom in the choice of reference matter type (then taken to elsewise have a 
negligibly small effect on the physics of the universe, so it is not observationally enforced) may embody 
a form of Multiple Choice Problem. 
On the other hand, demanding that such a reference fluid have further desirable properties (such as 
compliance with energy conditions in paragraph 1 or complete equivalence formulations in terms of 
quadratic constraint combinations with the usual physics in paragraph 4) might act as a filter on this.
It is then not clear if such a filter would be under-strong (so multiple inequivalent candidates 
remain), over-strong (so no candidates survive) or just right.

\section{Tempus Post Quantum}\label{SSec: TPQ}

{\bf Post Postulate} These are strategies in which time is not always present at the fundamental level. 
However, time is nevertheless capable of emerging in the quantum regime.  
Because this is an emergence, it means that the Hilbert space structure of the final quantum theory is 
capable of being (largely) unrelated to that of the WDE-type quantum theory that one 
starts with.   
Such emergent strategies are of the following types.

\subsection{Attempting a Klein-Gordon Interpretation based on Superspace Time} \label{SSSec: KGI}

The GR supermetric on superspace is indefinite like spacetime is, so analogous time-like notions exist 
for it. 
This suggests thinking about the WDE not as a time-independent Schr\"{o}dinger 
equation but as an analogue of the Klein--Gordon wave equation.  
%
%
In this way, this approach has QFT, and thus spacetime, undertones.
Some issues with this strategy are as follows. 

\mbox{ }

\noindent Problem 3.1-1.
Superspace null cones are not respected by superspace trajectories, limiting the analogy. 
%

\noindent Problem 3.1-2.  
There is an Inner Product Problem. 
Whether treated as a Tempus Post or a Tempus Ante scheme, the indefiniteness of the configuration space 
metric precludes such a scheme from having a Schr\"{o}dinger inner product (in parallel with 
Klein--Gordon theory).
The main issue then is whether it permits a Klein--Gordon inner product instead,\footnote{The capital 
Latin indices correspond to pairs of spatial indices and with the opposite raising/lowering convention, 
which jointly comprise the DeWitt 2-index to 1-index map \cite{DeWitt}.  
$\stackrel{\longleftrightarrow}{\delta}_B$ denotes functional derivative with both backwards as well as 
forwards action with respect to $h^{B} = h_{ab}$.} 
\beq
\langle\Psi_1[h]|\Psi_2[h]\rangle  = \frac{1}{2i}\prod\mbox{}_{\mbox{}_{\mbox{\scriptsize $x \in \Sigma$}}}
\int_{\mbox{\scriptsize Riem}(\Sigma)}d\Sigma_{A}G^{AB}(h)
\left\{
\Psi_1[h]\stackrel{\longleftrightarrow}{\delta_{B} }\Psi_2[h]
\right\} 
\mbox{ } , 
\eeq
However, there is a breakdown of the analogy between geometrodynamics and stationary-spacetime 
Klein--Gordon theory.  

\noindent
While there is a conformal Killing vector on superspace \cite{K81}, the GR potential does not in 
general respect this, and so this scheme fails.  
Strong gravity is a toy model that gets past this point, due to its much simpler potential happening to 
scale consistently. 
It is also an issue that the positive--negative modes split of states in the usual Klein--Gordon theory 
arises from the presence of a privileged time; thus, without such a privileged time in the general GR 
case, one's quantization scheme will not have this familiar and useful feature.    
%
%

\subsection{Third Quantization} \label{SSSec: 3rd}  

To get round the preceding SSec's problems, it has been suggested that the solutions $\Psi[h]$ of the 
WDE be turned into operators.  
This is analogous to the second quantization of a relativistic particle whose states are described by 
the Klein-Gordon equation (and thus this approach also has QFT undertones, and thus also spacetime 
undertones).
While Third Quantization is of interest as regards various technical issues, it was not held to provide 
a satisfactory approach to the Problem of Time up to the early 90's \cite{K92, I93}, and I am not aware 
of any subsequent advances in this respect.

\subsection{The Semiclassical Approach} \label{SSSec: SemiCl}

In the Semiclassical Approach \cite{DeWitt, LR79Banks, HallHaw, K92, I93, Kiefer9394, K99, Kiefer99, 
Kieferbook}, time is only meaningful in some semiclassical limit of the Quantum Gravity theory based on 
the WDE.   
This is e.g. a useful framework in non-primordial large-scale cosmology, to cover such as the origin 
\cite{HallHaw} of galaxies and cosmic microwave background perturbations.

In this approach, `heavy slow' (h) degrees of freedom provide a {\sl relative} time standard 
$t^{\sW\sK\sB}$ with respect to which `light fast' (l) local degrees of freedom run.  
Thus local physics has an emergent quantum dynamics based on (approximately) an emergent time-dependent 
Schr\"{o}dinger equation.   
E.g., in GR Quantum Cosmology, the h degrees of freedom are the size of the universe and homogeneous 
matter modes, and the l degrees of freedom are `shape' inhomogeneities in the universe's gravity and 
matter distribution.   
[Some modelling and possibly sometimes some truth in having gravitational field be h and matter be l. 
In the isotropic case, that is the same as scale.]  
For (e.g. using $h_{\mu\nu}$ as `h' and the matter as `l', and requiring that $|\chi\rangle$ depends 
nontrivially on h so that the QT is rendered nonseparable in h, l variables).
This approach may be viewed as a space-is-primary approach.  
One proceeds as follows.  
(I omit inclusion of ${\cal H}_i$ for simplicity).

By `heavy slow' and `light fast', what one means is that one uses 

\noindent 
1) the adiabatic Born--Oppenheimer-type ansatz 
\beq
\Psi = \psi(\mh^{A{\prime}})|\chi(\mh^{A^{\prime}}, \ml^{A{\prime\prime}})
\rangle
\label{WKBBO}
\eeq
and approximations of types often associated with this. 
[I use primed and double-primed indices for h and l.] 

\noindent 2) A WKB ansatz 
\beq
\psi(\mh^{A{\prime}}) = \mbox{exp}(iM\fF(\mh^{A{\prime}})) \mbox{ } .  
\label{WKB}
\eeq
(for $M$ a generic h-mass) and approximations of types often associated with that.    

\noindent Next, one forms the h-equation,  
\beq
\langle\chi| \hat{\cal H} \{|\chi\rangle\psi\} = 0 \mbox{ } ,  
\eeq  
which, under a number of simplifications and identifying $M\fF = \fW(\mh^{A{\prime}})$ Hamilton's 
principal function, yields 
\beq
N^{A{\prime}B{\prime}} \frac{\delta\fW}{\delta \mh^{A{\prime}}}
                         \frac{\delta\fW}{\delta \mh^{B{\prime}}} = 2\fU(\mh^{C{\prime}}) \mbox{ } 
\mbox{ } \mbox{ } \mbox{ (Hamilton--Jacobi equation)  .} 
\eeq
Here $\fU(\mh^{A{\prime}})$ is the h-part of $\fU$ and $N^{A{\prime}B{\prime}}$ is the inverse 
of the configuration space metric for the heavy degrees of freedom, $M_{A{\prime}B{\prime}}$.  
One way of solving this \cite{DeWitt, LR79Banks, HallHaw} is for an approximate emergent WKB 
semiclassical time $t^{\sW\sK\sB}(\mh^{A{\prime}})$, via the Hamilton--Jacobi theory relation plus 
momentum--velocity relation combination, 
\beq
\frac{\delta\fW}{\delta \mh^{A{\prime}}} = p_{\sh^{A{\prime}}} = 
M_{A{\prime}B{\prime}}\frac{\pa{\mh}^{B{\prime}}}{\pa t^{\sW\sK\sB}}  
\label{twe}
\eeq
(plus such as shift terms in general, ommitted by simplicity, and moreover, whose inclusion is 
unproblematic).
Furthermore, rearranging the subsequent equation makes it clear that $t^{\sW\sK\sB}$ is aligned with 
\cite{SemiclI} the form of $t^{\sJ\sB\sB}$ under the same regime of approximations.  
One also obtains a time-dependent Schr\"{o}dinger equation for the local l-degrees of freedom 
with respect to a time standard (approximately) provided by the background h degrees of freedom
\beq
i\delta|\chi\rangle/\delta t^{\sW\sK\sB} = \widehat{H}_{\sll}(x;l^{A\prime\prime}(x), t^{\sW\sK\sB}(x)]
|\chi\rangle \mbox{ } \label{TDSE2} \mbox{ } .
\eeq
This originates from rearranging a kind of fluctuation equation, 
\beq
\{1 - |\chi\rangle\langle\chi|\}\hat{\cal H}\{|\chi\rangle\psi\} = 0 \mbox{ } ,  
\eeq 
the emergent time dependent left-hand side term of (\ref{TDSE2}) arising from 
\beq
N^{A{\prime}B{\prime}} \frac{\delta^2\Psi}{    \delta \mh^{A{\prime}}\delta \mh^{B{\prime}}    } 
\mbox{ } \mbox{  containing the `crucial chroniferous cross-term' } \mbox{ } 
N^{A{\prime}B{\prime}} {i}\frac{    \delta \fW    }{    \delta \mh^{A{\prime}}    }
\frac{\delta|\chi\rangle}{\delta \mh^{B{\prime}}}   \mbox{ } ,
\label{CruCro}
\eeq 
which, via (\ref{twe}), $N^{A{\prime}B{\prime}}M_{A{\prime}C{\prime}} = 
\delta^{B{\prime}}\mbox{}_{C{\prime}}$ and the chain-rule gives

\noindent
\beq
i N^{A{\prime}B{\prime}}M_{A{\prime}C{\prime}}
\frac{\pa \mh^{C{\prime}}}{\pa t^{\sW\sK\sB}} 
\frac{\delta |\chi\rangle}{\delta \mh^{B{\prime}}} = 
i \frac{\pa \mh^{A{\prime}}}{\pa t^{\sW\sK\sB}} 
\frac{\delta |\chi\rangle}{\delta \mh^{A{\prime}}} = i \frac{\delta |\chi\rangle}{\delta t^{\sW\sK\sB}} 
\mbox{ } . 
\eeq
[$\widehat{H}_{\sll}$ is the remaining surviving piece of ${\cal H}$, which serves as Hamiltonian 
for the l-subsystem].  

\mbox{ }  

\noindent Problem 3.3-1.  Having invoked a WDE results in inheriting some of its problems \cite{K92, I93}.  

\noindent Problem 3.3-2.  Making the WKB approximation requires justification.  
Certainly (semi)classical conditions need not always occur \cite{I93}.  
These are quite a long list of conditions, and \cite{RPMCONCAT, SemiclIII, forth} show that a number of these 
do not occur in toy models.
The guarantee of a classical `large' as in the Copenhagen Interpretation of ordinary QT has been cast 
aside in considering the universe as a whole, and is then by no means be recovered in all 
possible situations.   
This justification gains substantially in importance upon realizing that witout the WKB ansatz 
\cite{Zeh86, BS, SemiclI}, one gets a complicated mess instead of an emergent time.  
See Sec 6 for one proposed resolution (decoherence). 
Moreover, even if this ansatz is justifiable in the non-primordial universe, say, that would only 
constitute a partial resolution of the Frozen Formalism Problem, as one would like to study the primordial 
universe too within the field of Quantum Gravity.  

\noindent Problem 3.3-3. It is furthermore unclear how to relate the probability interpretation of the approximation with that 
for the underlying WDE itself \cite{I93, K92}. 
And what of less approximate schemes, which are tied to better modelling of the back-raction of the 
light system on the heavy system? \cite{K92, SemiclI, SemiclIII, forth}.
(Some of these involve other than Schr\"{o}dinger equations.)

\noindent Problem 3.3-4. The status of the Spacetime Reconstruction Problem is unclear for 
the Semiclassical Approach \cite{I93}.

\noindent Problem 3.3-5. The Multiple Choice Problem remains present in such schemes in detail 
\cite{K92}.

\section{Type 1 Tempus Nihil Est}  

Adopting a Tempus Nihil Est approach  saves one from the thorny issue of trying to define time (going  
back at least as far back as Saint Augustin).
However, this is supplanted by two newer briar patches. 
1) how to explain the semblance of dynamics if the universe is timeless as a whole.
For, dynamics or history are in the present section to be {\sl apparent notions} to be constructed from the instant 
\cite{GMH, H99, HD, H03}.  

\mbox{ }

\noindent{\bf Type 1 Nihil Postulate}. 
One aims to supplant `becoming' with `being' at the primary level \cite{Page12, B94II, EOT, 
GMH, H99, HT, HD, H03} (see e.g. Reichenbach \cite{Reichenbach} for partial antecedents).   
In this sense, these Timeless Approaches are space-first ones.  

\mbox{ }

\noindent{\bf Nonstandard Interpretation of QT}.  
Type 1 and Histories Theory both soon become entwined in general questions about the 
interpretation of QT, in particular as regards whole-universe replacements for standard QT's 
Copenhagen Interpretation.
I note that some criticisms of Timeless Approaches \cite{K92, K99} are subject to wishing to preserve 
aspects of the Copenhagen Interpretation, which may in any case not be appropriate for Quantum 
Cosmology and Quantum Gravity.  

\mbox{ }

\noindent{\bf WDE Dilemma} Such approaches either 
(horn 1) invoke the WDE and so inherit some of its problems, 
or do not, thus risking the alternative problem 
(horn 2) of being incompatible with it, so that the action of the Wheeler--DeWitt operator kicks 
purported solutions out of the physical solution space.

\subsection{Na\"{\i}ve Schr\"{o}dinger Interpretation} \label{SSSec: NSI}

This interpretation of QT for the whole universe \cite{Hawking84, HP8688, UW89} is based on assuming 
that the inner product 
\beq
\langle\Psi_1[h]|\Psi_2[h]\rangle = \int_{\mbox{\scriptsize Riem}(\Sigma)}\Psi_1^*[h]\Psi_2[h] \d\Omega 
\mbox{ } .  
\label{NSIIP}
\eeq
gives the probability density for `finding' a spacelike hypersurface $\Sigma$ in $M$ with the 3-metric 
$h_{ij}$.
It can deal with issues of simple being, like ``what is the probability that the universe is large? 
Approximately isotropic? 
Approximately uniform (i.e. non-clumped, and so homogeneous)?'
One here addresses such questions via use of relative probabilities.  

\mbox{ }  

\noindent Problem 4.1-1.  However, it is conventionally regarded as of limited use since it does not 
permit answers to questions of ``being at a particular time", or of evolution/``becoming" \cite{K92}.    

\noindent Problem 4.1-2.  Time enters as an internal coordinate function of $h_{ij}$, and so is 
represented by an operator.  
However, as pointed out in \cite{I93}, there are problems with representing time as an operator. 

\noindent Problem 4.1-3.  This approach is menaced by horn 2 of the WDE Dilemma 
via its thus-named `na\"{\i}ve' inner product postulation.

\subsection{Conditional Probabilities Interpretation} \label{SSSec: CPI}

Conceptually, this is a refinement of the na\"{\i}ve Schr\"{o}dinger interpretation that extends the 
range of questions it can answer

\noindent (thus it is an improvement as regards Problem 4.1-1); technically and 
as interpretations of QM, the two schemes are highly distinct.      
It was proposed by Page and Wootters \cite{PW83} (see also the comments, criticisms and variants in 
\cite{K92, K99, Giddings, Pullin}.)   
It involves considering conditional probabilities for the results of a pair of observables $A$ and $B$, 
including outside the context of a pre-existing notion of time, so that this is a base object that is 
being postulated to be meaningful in a way that is outside the conventional formalism of QT.

As this base object involves a pair of observables $A$ and $B$, postulating it to be a 
meaningful quantity has the consequence that answers to questions of `being' that concern correlations 
between $A$ and $B$.  
Such a question might be ``what is the probability that the universe is almost-flat {\sl given} that it 
is almost-isotropic?"
One addresses such questions by via postulating the relevance of conditional probabilities
\beq
\mbox{Prob}(B\in b|A\in a, \rho) = \frac{\mbox{Tr}
\big(
P^B_b\,P^A_a\,\rho\,P^A_a
\big)                            }{
\mbox{Tr}
\big(
P^A_a\,\rho
\big)
} \mbox{ } ,      
\label{Pr:BArho}
\eeq
for finding $B$ in the subset $b$, given that $A$ lies in the subset $a$ for a system in state $\rho$. 
[Here, the projection operator for an observable $A$ an observable and $a$ a subset of 
the values that this can take is denoted by $P^A_a$.]

Moreover, such a scheme has been suggested to be general enough to cover all types of questions that 
occur in science. 
Firstly, one may then hope that an $A$ can be found that is a sufficiently good clock variable, so that 
the above question about $A$ and $B$ can be rephrased to concern what value $B$ takes when the clock 
variable $A$ indicates a particular time \cite{PW83, I93}.  

\mbox{ }

\noindent Problem 4.2-1. {\sl Good} clock variables are far from guaranteed to be available.

\noindent Problem 4.2-2. 
This is as far as the original Conditional Probabilities Interpretation goes; it is not a full 
resolution of Problem 4.1-1 as it does not per se address questions of becoming; for a further 
extension to address those too, see the ``Records A)" scheme in the next SSec.

\noindent Problem 4.2-3. Horn 1 of the WDE Dilemma applies here \cite{K92}.  

\mbox{ } 

\noindent 
Finally, there is also a Pullin--Gambini--Porto \cite{Pullin} formulation that also uses conditional 
probabilities; this Formulation gives a modification of the Heisenberg equations of motion of the 
Lindblad type, which is tied to the incorporation of decoherence.

\subsection{Records Theory Schemes} \label{SSSec: Records}

These are further Timeless Approaches, in which what is considered as primary are {\it records}, i.e. 
some sort of instantaneous configuration.     
One then seeks to construct a semblance of dynamics or history from the correlations between these  
records.
Theoreticians have, however, differed somewhat in how to make the notion of record more precise, and in 
how they envisage the semblance of dynamics may come about.    
Thus there are in fact a number of Records approaches.
I caution that, while passing to scheme D) removes some of the problems of schemes A) and B) [and of 
histories-theoretic elements from scheme C)], this remains incomplete.  

\mbox{ } 

\noindent Records Scheme A) Page \cite{PAOT, Page12} has further advanced the Conditional Probabilities  
Interpretation, by considering in principle that present events contain memories of the past.  
Thus one could now have correlations being between `records of past events', from which to extract  
predictions about `future events'.
%
%
One might view such configurations as e.g. researchers with data sets who remember how they set up the 
experiment that the data came from (controlled initial conditions, and so on).
Unfortunately, this is very speculative from the perspective of doing concrete calculations.  
It would be expected to be very hard even to toy-model, one would need a working IGUS model (c.f. 
Appendix A.9).
One benefit of such an approach is that a concrete conditional probability object is to be computed 
and interpreted [this is more specific than Records Scheme B)].

\mbox{ } 

\noindent Records Scheme B) The Bell--Barbour approach \cite{Bell,B94II,EOT} (see \cite{ButterBar} for 
further differences between these authors) reinterpret Mott's calculation \cite{Mott} of how 
$\alpha$-particle tracks form in a bubble chamber as a ``time capsules" paradigm for Records Theory.  
Barbour has then argued for Quantum Cosmology to be studied analogously, with somewhat similar 
arguments being made by Halliwell \cite{H99, HD, H03} and by Castagnino--Laura \cite{CaLa}.
Barbour's own approach additionally involves 1) reformulating classical physics in timeless terms 
\cite{BB82, B94I, RWR, EOT}.  
[N.B. that \cite{EOT} should not be read to be literally placing emphasis on timelessness casting 
mystery upon why `ordinary physics' works.]  
This leads to the JBB time of Sec 2.2, and thus perhaps contact should be made with the Semiclassical 
Approach's WKB time.  
2) He follows Leibniz in emphasizing the configuration of the universe as a whole, and how this is the 
only perfect clock (and then takes JBB time to embody this whole-universe character).  
However, I consider this whole-universe configuration feature to make the scheme impractical, as we do 
not have that detailed a knowledge of the universe as a whole.  
3) He furthermore speculates \cite{B94I,B94II,EOT} that the asymmetry of the underlying curved 
stratified quotient configuration space 
causes concentration of probability density on ``time capsules" rather than other instantaneous 
configurations.  
However, it is not yet clear whether 3) is qualitatively or quantitatively vindicated; RPM models 
may help with this issue (as a preliminary result, my first few simple RPM models \cite{RPMCONCAT} would 
not seem to exhibit any substantial peaking about such configurations.)

\mbox{ }
 
\noindent Records Scheme C) Gell-Mann--Hartle \cite{GMH} and Halliwell \cite{H99, HT, H03} have found 
and studied records contained within Histories Theory (see two SSecs down).  
Moreover, viewpoints that presume and derive histories are logically distinct, with 
the latter 

\noindent requiring more work than the former. 

\mbox{ }

\noindent Records Scheme D) My own approach to Records Theory \cite{Records, forth} does not follow any 
one of A), B) or C) above, but is rather a synthesis of elements drawn from each of them.  
I consider records to obey the following 2 postulates.  

\mbox{ }

\noindent {\bf Records Postulate 1}. Records are information-containing subconfigurations of a single 
instant that are localized in both space and configuration space.\footnote{Here
Barbour's whole-universe requirement 2) of scheme B) is not adopted due to the above critique of it.}
%
\noindent Local in space means `under one's nose'.
This is partly so that they are controllable, and partly so that one has more than one such to compare.  
It also negates signal times within conventional frameworks in which such are relevant (however, this 
is far from necessarily the basis for a criticism; e.g. p 225/226 of \cite{Whitrow} points out that 
relativistic theories make use of a similar notion of locality).
\noindent Local in configuration space has to do with imperfect knowledge and different levels of 
imperfection of knowledge, i.e. fine and coarse grainings.\footnote{This follows from making a good  
case, via C), that Records Theory should possess counterparts of those history-theoretic structures 
(see Sec 5) that remain meaningful at the kinematical level.}   

\mbox{ }

\noindent {\bf Records Postulate 2}. Records are furthermore required to contain useful information. 
I take this to mean information that is firstly and straightforwardly about correlations. 
Secondly, however, one would wish for such correlation information to form a basis for a semblance of 
dynamics or history.  
I have a hunch here that one will be disappointed (Possible Problem 4.3-1). 
This follows from there being two paradigms for records, not only the Mott--Bell--Barbour bubble 
chamber track, but also the Joos--Zeh \cite{JZ} dust grain being decohered by microwave background 
photons.  
The fear is then that the latter is much less contrived and thus likely to be much more common in the 
universe, so that one would seldom be able to deduce much from records. 
Some hope might return to the scheme via Halliwell's finding usefulness even in very imperfect records 
(see Sec 6).
This is crucial as to the sensibleness of using few-particle RPM's for records-theoretic investigations. 
These have as much or more capacity to form records as Halliwell's model (whilst also now being 
understandable as whole-universe models of such a process).  

\mbox{ }

\noindent
Such a scheme then requires \cite{Records, forth} suitable notions of distance in configuration space, 
locality in space, information, relative information and correlation.   

\mbox{ }  

\noindent Problem 4.3-2 (technical incompleteness).
Many of these are difficult to provide for general GR.  
[But RPM's are rather more amenable in these respects, making them good toy models for studying Records 
Theory.]  

\mbox{ } 

\noindent Moreover, even possessing a compatible such set of notions does not guarantee a mechanism by 
which a semblance dynamics or history arises.  

\mbox{ }  

\noindent {\bf Records Theory} is the subsequent study of how dynamics (or history or science) 
is to be abstracted from correlations between such same-instant subconfiguration records.  
%

\mbox{ }  

\noindent Problem 4.3-3 (how to make this into a fully fledged strategy is still the subject of 
speculation). In scheme D), I make as yet no commitment to any of Barbour or Page's undemonstrated   
mechanisms for acquiring a semblance of dynamics (Page's one serves in principle but is of a 
highly impractical form).  
I view RPM's, detector models and IGUS'es (and as-yet hypothetical combinations of such) as useful toy 
models for investigating this most mysterious aspect of Records Theory. 
It may also be necessary to resort to assuming e.g. Histories Theory, and/or elements of the 
Semiclassical Approach (Sec 6), in order to obtain a robust mechanism for obtaining a semblance of 
dynamics from records.

\section{Non Tempus Sed Historia}  

{\bf Histories postulate}. Supplant working in a canonical fashion with configurations to treating 
histories as primary.  

\mbox{ } 

\noindent One implementation of this starts from Feynman path integrals, 
\beq
\Psi[Q] = \int {\cal D}Q \, \mbox{exp}(-i\fS[Q]) \mbox{ } ;     
\eeq
moreover, it then builds up much further structure. 
Individual histories are built using strings of projectors $P^{A_i}_{a_i}$, i = 1 to $N$ 
\beq
H := P^{A_N}_{a_N}(t_N) . . . P^{A_1}_{a_1}(t_1) \mbox{ } .  
\eeq 
Assigning probabilities to such does not work in QT.  
For, if $a(t)$ has amplitude $A[a] = \mbox{exp}(i\fS[a])$ and $b(t)$ has amplitude $B[b] = 
\mbox{exp}(i\fS[b])$, then these are nonadditive since in general $|A[a] + B[b]|^2 \neq |A[a]|^2 
+ |B[b]|^2$ in general.
In fact, Histories Theory involves \cite{GMH, Hartle} a far-reaching \cite{I93} extension of normal 
Quantum Theory to a form outside its conventional Copenhagen interpretation: it is a many-worlds scheme.

The above description provides some arguments \cite{I93} towards regarding Histories Theory as an 
expansion of the Type-1 

\noindent Tempus Nihil Est approaches.  
However, Histories theory also has connotations of being a spacetime-first approach from the 
philosophical perspective.  
It shares further technical features with the Internal Time Approach, to the extent that one can argue 
that it developed due to the failings of the Internal Time Approach.

A further ingredient for Histories Theory is the notion of fine/coarse graining. 
Additionally, one requires a precise prescription from within the formalism itself that says under 
which circumstances it is meaningful to ascribe a probability to a history. 
Toy models used for this scheme include ordinary and relativistic mechanics and electromagnetism 
\cite{Hartle}.
In the 

\noindent extension to the GR case, histories are defined such that no direct reference to time is needed.
The final traditional ingredient of the Histories Theory approach is the {\it decoherence functional} 
between a pair of histories $H$, $H^{\prime}$,
\beq
D(H^{\prime}, H) = \mbox{tr}(H^{\prime}\rho H) \mbox{ } ,
\eeq 
This is useful as a `measure' of interference between $H^{\prime}$ and $H$.  
It is zero for perfectly consistent theories.  
It has Hermitian, positive, normalized and superposition properties.

\subsection{Other more or less conventional Histories Approaches}\label{SSSec: +Hist}

Histories Theory has been further developed by Isham and Linden \cite{IL} as the {\it Histories 
Projection Operator} or {\it Generally-Covariant Histories Formalism}. 
This identifies the maths that Histories Theory involves. 
It is an orthoalgebra, or, less minimalistically, a lattice of propositions.\footnote{There is a (simpler) 
Records Theory counterpart of this, since these operations remain meaningful in that context.}

The Histories Projection Operator Approach (see also \cite{ILSS, Savvidou1b, SavRest}) has two notions 
of time that are considered to be distinct. 
On the one hand, there is a kinematical notion of time that labels the histories as sequences of events  
(the `labelling parameter of temporal logic').
On the other and, there is a dynamical notion of time that is generated by the Hamiltonian.  
In the Histories Projection Operator Approach (but not in the previous SSec's schemes), histories are 
themselves realized as projection operators in some Hilbert space. 
(In \cite{GMH}, histories are the product of Heisenberg picture projection operators, which product is 
usually not itself a projection operator).

In this kind of approach, the histories postulate is implemented by still considering a canonical 
approach, but whose underlying objects are not configurations $q^a$ but histories $q^a(t)$ (for t now a 
continuous label).  
These then are to have canonical momenta $p_a(t)$ and one is furthermore to use Poisson brackets  
(`histories brackets') 
$$
\{q^a(t),p_b(t^{\prime})\} = \delta^a\mbox{}_b\delta(t,t^{\prime}) \mbox{ } .  
$$ 
Kouletsis and Kucha\v{r} \cite{KoulKu, Kouletsis08}'s classical work on GR, and on the bosonic string 
as a toy model, involves a {\it space map} as well as a {\it time map} for how the family of geometries 
along each history embed into spacetime; this work also makes substantial contact with the Internal 
Time Approach and with the Problem of Observables.
Then, one goes about quantization (only carried out to date for simpler models: ordinary mechanics 
\cite{IL, ILSS}, relativistic mechanics \cite{Savvidou1b}, minisuperspace \cite{AS05} and quantum field 
theories \cite{Anastopoulos}) via replacing the canonical group by a histories group that is to play a 
similar role. 
One takes time to be continuous in such approaches, so that one obtains a 1-$d$ QFT in time (even for 
finite toy model examples).  

\mbox{ }  

\noindent Question. 
What happens to the more complicated cases \cite{KoulKu, Kouletsis08} of this scheme at the quantum 
level is not as yet known.

Marolf \cite{Marolf} has a distinct way of getting history brackets: he uses the Hamiltonian as an 
extra structure to extend the Poisson bracket from a Lie bracket on phase space to a Lie bracket on the 
space of histories.  
In the usual Histories Brackets Approach, however, one puts the equal-time formalism aside here and 
introduces a new phase space in which the Poisson bracket is defined over the space of histories from 
the beginning.  

\mbox{ } 

\noindent Problems 5-1 to 3. The preceding SSec's Histories Theory has the problem that the path integral form 
of the decoherence functional is hard to define mathematically (and looks prone to Multiple Choice 
and Functional Evolution issues \cite{K92}). 

\noindent 
The current SSec's Histories Theory provides computational tools of some use at least toward the first 
of these issues. 

\noindent Problem 5.1-1.
The current SSec's Histories Theory is, however, far from finished work for full GR; while Kouletsis 
looks to have good control of the classical parts of the Problem of Time for full GR, the quantum 
counterpart of such work has not as yet appeared (not even for the bosonic string toy model).  

\mbox{ }

\noindent 
Sorkin's Causal Sets Approach \cite{Sorkin03} could be viewed as much like a Histories Theory in which 
less structure is assumed.  
Here, causal ordering as occurs in classical spacetime is taken to be the fundamental notion to be kept 
in Quantum Gravity. 
However, Spacetime Reconstruction difficulties follow from this approach's insistence on very sparse 
structure.

\section{Suggested composites of Semiclassical, Records and Histories ideas} \label{Sec: Collate}

Considering these together makes sense insofar as there are indications that these are able to prop 
each other up toward providing a more robust candidate strategy for resolving the Problem of Time 
and addressing various other fundamental issues in Quantum Cosmology \cite{H03}.  
These indications are as follows. 

\noindent 1) Histories Theory contains records, while Records Theory hopes to abstract history from 
records (this is the aforementioned work by Gell-Mann--Hartle, and by Halliwell and collaborators.)  
Records Theory in this approach benefits by inheriting part of the structural framework that has been 
developed for Histories Theory (\cite{Records, forth}, Sec 4.3).

\noindent 2) Decoherence of histories is one possible way of obtaining a semiclassical regime in the 
first place (e.g. \cite{H09, Kiefer09} have argued in favour of decoherence as the mechanism that 
turns such more general superposed states into WKB states).  
 
\noindent 3) What the records are will answer the further elusive question of which degrees of freedom  
decohere which others in Quantum Cosmology. 
As Gell-Mann and Hartle say \cite{GMH}, records are the ``{\it somewhere in the universe that 
information is stored when histories decohere}".

Halliwell \cite{H99, HT, HW} has made subsequent advances with 1) to 3). 
E.g. he has found that even small models are capable of forming records that are now imperfect. 
(They cannot store substantial information about interactions, but can store at least some nontrivial 
information about interactions, e.g. a detector model that has one degree of freedom, see Appendix A.8). 
He makes use of Wigner functions (see e.g. \cite{HT}) and of Semiclassical Approach ideas 
(see e.g. \cite{H09}).  
These works mostly involve fairly standard models, with known wavefunctions allowing for computation of further 
useful objects \cite{H99} such as the Feynman--Vernon influence functional. 
These systems have to date largely not had whole universe model connotations. 
These can be included without sacrificing computability by using RPM's (though this remains work in 
progress of mine), which work can span all of Histories Theory, Records Theory and the Semiclassical 
Approach, making RPM's a good prospective arena in which to attempt to unite these.

\section{Type 2 `Rovelli' Tempus Nihil Est}

These began with \cite{Rov91a, Rov91b, Rov91c} (though one might view \cite{DeWitt62W86, PW83} as 
forerunners in some ways, see also \cite{Car9091b}).
They involve various types of observables and related concepts, as defined below.

An {\it observable} or {\it constant of the motion} or {\it perennial} in classical canonical 
GR is any functional ${\cal O}[Q,P]$ of the canonical variables whose Poisson brackets with all the 
constraint functions vanish (perhaps weakly \cite{I93}).  
Thus, for geometrodynamics  
\beq
\{{\cal O}, {\cal H}_{i}(x)\} = 0 \mbox{ } ,                     
\label{OHi0}
\eeq
\beq
\{{\cal O}, {\cal H}(x)\} =  0 \mbox{ } .                   
\label{OH0}  
\eeq
(This is a concept that goes back to Dirac and Bergmann).
Justification of the name `constants of the motion' follows from the Hamiltonian for GR 
being $H[\,N, N^{i}\,] := \int_{\Sigma} \d^3x\,(N{\cal H} + N^{i}{\cal H}_i)$ 
so that (\ref{OHi0},\ref{OH0}) imply that
\beq
\frac{d{\cal O}}{dt}[h(t),\pi(t)] = 0 \mbox{ } .
\ee
Thus, observables are automatically constants of the motion with respect to evolution along 
the foliation associated with any choice of $N$ and $N^i$. 
The quantum counterpart of these then involves some operator form for the canonical variables and 
commutators in place of Poisson brackets.  
The operator-and-commutator counterparts of the above are then another manifestation of the Frozen 
Formalism Problem of classical canonical GR (some sort of `Heisenberg' counterpart of the 
`Schr\"{o}dinger' WDE being frozen).

\noindent However, some Quantum Gravity practicioners consider that (\ref{OHi0}) alone may suffice, and 
thus drop requirement of (\ref{OH0}); this defines a {\it Kucha\v{r} observable} (see also \cite{BF08}).  

\noindent 
A {\it true observable} (Rovelli 1991) alias (in 2002) {\it complete observable} also termed (at least 
in \cite{Thiemann}) {\it evolving constant of the motion} is an operation on a system that produces a 
number that can be predicted (or whose probability distribution is predicted) if the (in the case of QT, 
Heisenberg) state of the system is known.  

\noindent This is as opposed to a {\it partial observable} (Rovelli 1991), which is an operation on 
the system that produces a number, the number possibly being totally unpredictable even if the state is 
perfectly known.  

\noindent Note that all of the above definitions were, more or less, in place by 1991, though then early 
90's and 2000's approaches do in part differ as to which of these definitions they use and how these 
definitions enter into creating strategies toward resolving the Problem of Time.

\subsection{Early 90's forms}

Rovelli argued that the `evolving constants of motion' form a particularly important class by which some 
notion of dynamical evolution can be introduced.
The case of a simple model with just one constraint serves well to explain how such a scheme would work.  
Here one has a toy model Hamiltonian constraint $H(q,p)$ defined on a finite-dimensional phase 
space ${\cal S}$.  
One then introduces some internal time $T(q, p)$ such that $\forall \mbox{ } t \in 
\mathbb{R}$, the hypersurface
\beq
{\cal S}_t =  \{(q,p)\in{\cal S}|T(q,p) = t\}
\eeq
intersects each dynamical trajectory (on the constraint surface) generated by $H$ exactly once. 
Thus $\{ T, H\} \neq 0$, and so the internal time $T$ itself is not a physical observable in the above 
sense.
One then associates with each function $F(q, p)$ a family of observables (`constants of motion') $F_t$, 
$t\in \mathbb{R}$, defined by  
\beq
\{F_t, H\} = 0  \mbox{ } \mbox{ } , \mbox{ } \mbox{ }  F_t|_{{\cal S}_t} = F|_{{\cal S}_t}     
\label{12} \mbox{ } . 
\eeq
Thus, the observable $F_t = F$ on the subspace ${\cal S}_t$ of ${\cal S}$. 
Dynamical evolution with respect to $T$ then amounts to how the family of observables 
$\{F_t | t\in \mathbb{R}\}$ depends on $t$.  
In this way, this scheme is \cite{Rov91a} a classical analogue of the Heisenberg picture of QT.
Such a model could e.g. comprise a classical mechanics (or minisuperspace) model for which the role of 
time variable 

\noindent has been taken over by an internal time $T$.  

Rovelli then suggested that these evolving constants of motion should form the basis for a 
Heisenberg-style quantization 

\noindent of the system.  
Moreover, Rovelli further argued that (see also e.g. \cite{RovelliPMP}) for Quantum Gravity such a 
Heisenberg picture scheme remains meaningful whilst the Schr\"{o}dinger picture no longer is.  
Two ways of applying these ideas toward building Quantum Gravity programs are as follows \cite{I93}.  

\mbox{ }

\noindent Implementation 1) Use the group-theoretical scheme in \cite{I84}, and hope that 
this produces a Lie algebra of the $F_t$, allowing one to then find a self-adjoint operator 
representation (`the quantum observables') from obtaining unitary representations of the associated 
Lie group.  
Thus one would construct a physical Hilbert space, giving one a probabilistic interpretation without 
having made any specific identification of time.

\mbox{ }  

\noindent Implementation 2) Start with a Tempus Post Quantum scheme that uses an operator 
representation of the canonical commutation relations on some Hilbert space ${\cal V}$.
Then physical state space ${\cal V}_{\rm phys}$ consists of all vectors in ${\cal V}$ that satisfy the 
operator constraints (\ref{calH}, \ref{calHi}).
Then a physical observable is defined to be any operator $\widehat{\cal O}[\widehat{h},\widehat{\pi}]$ 
that satisfies the commutator of operators analogue of (\ref{OHi0}, \ref{OH0})  
%
%
[or perhaps a weaker version in which these vanish only on the physical subspace.]

\mbox{ } 

\noindent{Some problems with the early 90's forms are as follows.}

\noindent Problem 7.1-1. A common problem raised against Type 2 `Rovelli' timeless schemes (which applies to both of 
the above implementations \cite{I93} and to the subsequent SSec's, concerns not knowing how in general 
a sufficient set of concrete such physical observables are actually to be constructed.  
Observables may be extremely rare and hard to find. 
(In particular, Kucha\v{r} likens perennials to unicorns in \cite{K93}, as both have exceedingly 
wonderful properties but are also exceedingly unlikely to be available for one to use.   
This is a large part of why Kucha\v{r} has advocated his definition of a weaker kind of observable. 
Even at the classical level, satisfying (\ref{OHi0}, \ref{OH0}) requires solving the classical 
equations of motion, which, in a sufficiently general setting, is a very tall order; Rovelli's early 
papers and H\'{a}j\'{\i}\v{c}ek agree on this point).    
Moreover, Torre and Anderson \cite{AT93} showed that observables for GR are necessarily highly nonlocal. 

\noindent Problem 7.1-2. Furthermore, Implementation 1 is a type of program for which sets of observables 
that generate a true Lie algebra are rare. 
It is more likely for the algebra produced to be one not with structure constants but with nontrivial 
structure functions of the canonical variables.  
Ordering ambiguities then make it difficult to find self-adjoint representations of such algebras 
\cite{Haj91, K92, I93}. 
This is in particular due to difficulties with promoting to operator form the first condition in 
(\ref{12}) [in practise, this is likely to be an implicit function of $F$, which entails the kind of 
problems in SSec \ref{SSec: ISI}].

 \noindent Problem 7.1-3. Implementation 2 then additionally encounters an Inner Product Problem 
\cite{I93}.

\noindent Problem 7.1-4. As the time function $T$ invoked is an internal time, it is prone to global 
obstructions as per Sec 2.  
One may hope again that $T$ being a matter time rather than geometrodynamical could circumvent 
this in a satisfactory manner.  

\noindent Problem 7.1-5. Hartle \cite{Hartle96} has further objected on grounds of the unlikeliness that 
a single Hilbert space can be used for all possible choices of internal time function $T$. 

\noindent Problem 7.1-6. Hartle furthermore argues that this approach has a Functional Evolution Problem, 
and that inconsistent time orderings arise within it.

\subsection{Forms from the 2000's}

More recently, further suggestions have been made as regards extracting Problem of Time strategies from 
timeless notions of observables.   
The Partial Observables Approach \cite{035} involves quantities that are not predictable but the 
correlations between which are. 
Rovelli then argues that QT should be about the relations between partial observables. 
Thiemann \cite{Thiemann01} argues even more generally that only evaluation of one thing relative to 
another will be meaningful.
[The first of these argued points I view as likely, and the second as very almost certain. 
However, I point out that such arguments are unlikely to be confined solely to Type 2 `Rovelli' Tempus 
Nihil Est Approaches.
E.g. similar ideas are present in the Conditional Probabilities 
Interpretation/Records Theory programmes; see Sec 8.1 for further comparison of these strategies.]

Thiemann \cite{Thiemann} gives the following argument (see also Rovelli \cite{Rovellibook, 035} for 
earlier work in this direction).  
Let $A$, $B$ be two non-observables that are, however, partial observables.  
Then these are not then invariant under the gauge flow of the Hamiltonian constraints. 
Let their flows with respect to one of the Hamiltonians be $\alpha_t(A)$ and $\alpha_t(B)$.  
Then fix some parameter $\tau$ and invert the condition $\alpha_t(B) = \tau$ for t.  
Next, insert the solution $t_B(\tau)$ into the function $t \mapsto \alpha_t(A)$ and obtain $A_B(\tau)$.  
Then $A_B(\tau)$ is invariant under $\alpha_t(A_B(\tau)) = A_B(\tau)$ and is interpreted as the value 
of $A$ when $B$ has the value $\tau$.  
Thus, while neither $A$ or $B$ are invariant, we can construct a simple invariant $A_B(\tau)$ which is 
frozen but still has a dynamical interpretation.
This, Thiemann argues, is ``precisely how we perceive time in physics" \cite{Thiemann} ($B$ playing role 
of a time, $T$), all that's actually observable being relative motions.
Thiemann then claims that $\tau$ moreover can be shown to have a canonical generator, i.e. a physical 
Hamiltonian, which is completely independent from the Hamiltonian constraint, itself being gauge 
invariant and not having to vanish on the constraint surface.  
Parts of this scheme do have some conceptual similarities to the Conditional Probabilities 
Interpretation; note however that it is not tied to the specific Conditional Probabilities formula for 
extracting information about the correlations.

The above is a line of work that Pons and Salisbury \cite{0503013} have supported.  
It is also used by a number of authors working toward Loop Quantum Gravity. 
That GR has an infinity of Hamiltonian constraints rather than just the one -- time is a function of 
$x^{\mu}$ rather than a single number -- causes a number of complications, can also be circumvented by 
Thiemann's Master Constraint Program \cite{Phoenix, Thiemann} in the Ashtekar Variables setting (see 
SSec 8.3 for more about this).

\mbox{ }  

\noindent
Dittrich's alternative line of work \cite{Ditt} has extended the early 90's formulation and tied it to 
Kucha\v{r}'s `bubble time' Internal Time scheme (whence a number of global issues become apparent: 
problem 2.5-3). 
This work is at the classical level.  
It does not contain many theory-dependent specifics, though it does include cylindrical waves 
midisuperspace as an example for which the scheme does work out (as one might expect from the tie to 
Internal Time schemes).  

\mbox{ }   

\noindent{Some problems with the 2000's forms are as follows.} 
%
I begin by noting that the Partial and Complete Observables concepts in terms of which the 
newer work is phrased have themselves already been part of Rovelli's program in 1991. 
Thus, the presence of these concepts by themselves does not constitute a valid argument as to the 
inapplicability of the existing critiques to the newer work. 
The issue is, rather, on how these concepts are used in the building of new Problem of Time strategies that 
attention should be concentrated. 
By the 2000's work being the early 90's work plus additional ideas, a number of the critiques of the 
early 90's work would be expected to carry over to the 2000's work.  
[How these strategies might get round the preceding criticisms whilst hopefully not generating any new 
criticisms, has not yet (as far as I know) been covered in detail in the literature.  
Whilst I gather and form some such points below, I do not as yet have a definitive detailed critique, 
but merely pose that as a project that would be of current interest.]
Some particular issues are as follows.  


\noindent Problem 7.1-1  -- the Problem of Observables that finding full sets of explicit 
observables is a tough and as yet unanswered question -- continues to be acknowledged in the more 
modern work of Thiemann \cite{Thiemann} and Dittrich \cite{Ditt}. 
Though I do point out that suggestions representing at least some progress \cite{0111, Rovelli02, 
Smolin08, otherPOObs} have been made, albeit I do not have the space here to present and analyze this 
work in detail.  
I also point out that the increase in emphasis on partial observables deals with this problem 
since partial observables themselves are not problematic to find.

\noindent Problem 7.2-1: How does the above partial observables approach square with the 
well-established theoretical issue (c.f. page 4) that some clocks/timefunctions are better than others?     
%
%
Thiemann \cite{Thiemann} admits that it may not be easy to find suitable clock variables.     
Rovelli furthermore argues in e.g. \cite{RovelliPMP} for relative motion of variables to be a 
democratic issue in the sense that any $A$ can serve as a timestandard for any $B$, rather than there 
being some privileged subset of reliable timestandards.
Moreover, he uses Newtonian mechanics as an example of such a scheme.  
However, this particular example does not take into account \cite{Zeh09} that Newtonian mechanics makes use not of democratic 
timestandards but rather of a meritocratic timestandard: the Newtonian time is used since it exhibits 
the advantageous property that the equations of motion are particularly simple with respect to it.  
Nevertheless, the democracy argument is probably in fact intended for generic GR. 
A further issue is that if one does argue to use any $A$ as a timestandard for any $B$, then one may 
stand to lose the distinction between observables and purported timestandards at the level of quantum 
operators that permits such purported timestandards to indeed turn out to be good timestandards at the 
quantum level.    

\noindent Possible Problem 7.2-2.  Might the inversion in the second part of this SSec be problematic in 
general to explicitly perform?

\noindent Problem 7.2-3. Thiemann \cite{Thiemann} and Dittrich \cite{Ditt} acknowledge that finding 
observables (not taken to include partial observables) is further complicated by global issues (which amount to a Global Problem of Time); this remains 
particularly clear due to the connections with Internal Time Approaches persisting from the early 90's 
schemes into 2000's schemes.   


\noindent Moreover, Rovelli \cite{Rovelli02} argues that the location with respect to the matter 
in models with matter makes the partial observables approach more straightforward.  


\noindent Problem 7.2-4. However (in many ways similarly to Problem 2.6-2), it is not clear whether 
reference fluids/clouds of clocks meaningfully exist.  

 
\noindent On the other hand, using a such could (in parallel with Sec 2.6) remove the Global Problem.

\section{Conclusion and Discussion}

\subsection{Summary and some overarching comments}

`Time' means incompatible things in each of general relativity (GR) and quantum theory (QT), which is 
problematic in trying to put these two theories together to form a theory of Quantum Gravity. 
That time goes missing in attempting to write down a canonical formulation for Quantum Gravity evokes 
the various strategies of 

\noindent I) finding time before quantization (Tempus Ante Quantum). 

\noindent II) Finding time emerge after quantization under certain circumstances (Tempus Post Quantum).  

\noindent III) Dealing with as much physics as is possible from a timeless perspective. 

\noindent IV) Ditching the notion of time for the notion of history instead, and re-starting the 
canonical quantization process using these instead of configurations.  
Moreover, the above `Frozen Formalism Problem' is but one of a number of facets of the Problem of Time, 
which I have argued form a coherent whole (`Ice Dragon') due to stemming jointly from the crisis that 
GR and QT have relied upon mutually incompatible notions of `time'.

The above problems, and strategies were subjected to detailed critical review in the early 90's by 
Kucha\v{r} and Isham \cite{K91, K92, I93, K93}.  The present article covers a number of subsequent 
advances in the strategies. 
In particular, 

\mbox{ } 

\noindent
1) a wider range of Histories Theories have appeared.


\noindent
2) The division between Rovelli, and other previously proposed, timeless strategies has continued to 
grow, with each side of this divide producing fruitful new work. 
Namely, on the one side, Rovelli's Partial Observables Approach, Dittrich's work and the application of 
some such ideas to Loop Quantum Gravity, in particular by Thiemann (see below).  
On the other side, Page, Barbour and I have made a number of proposals involving Timeless Records, with 
records also being found and studied within Histories Theory by Gell-Mann and Hartle, and Halliwell.
As some useful contrast, there are some parallels between `Rovelli'-type approaches and 
Page--Wootters' original Conditional Probabilities Interpretation \cite{PW83}.  
Both involve notions of observables, and of subsequent pairings of such, with the one then serving as 
a clock for the other.
However, Rovelli's scheme is less specific in not postulating a rigid formula like the Conditional 
Probabilities Interpretation one for the extraction of the information about the correlation between the 
two observables.  
Also, Rovelli insists that only the Heisenberg picture remains meaningful for Quantum Gravity, by which 
Type 1 Approaches such as the Conditional Probabilities Interpretation/Records Theory could not be just 
the same physics as Type 2 `Rovelli' but in a different picture.
(Though it is not clear, at least to me, whether this insistence is absolutely necessary; attempting to 
lift it might reveal further approaches to the Problem of Time, or further connections between existing 
approaches, such as between Type 2 `Rovelli' and Type 1 Timeless Approaches.  
Moreover, Isham and Dittrich established connections between Type 2 `Rovelli' Approaches and Internal 
Time Approaches, 
%
%
which do constitute at least an emergent Schr\"{o}dinger picture.) 
Also, Type 1 involves nonstandard interpretation of QM; is this also so of Type 2 `Rovelli'? 
(Does the Heisenberg picture alone remaining meaningful entail some features of this?
Thiemann's combination of Type 2 `Rovelli' with Histories Theory certainly involves bringing in 
nonstandardness of interpretation of QT.)  
Finally, cautions about requiring the observables cast in the role of clocks/timefunctions to actually 
be (often hard to find) {\sl good} clocks/timefunctions should then apply to both Type 1 and Type 2 
schemes.  

\mbox{ }

\noindent 3) Comparing and composing Problem of Time strategies has started to appear in the literature.  
Examples of this are 

\noindent
i) Halliwell's \cite{H99, H03, HD, HT, HW, H09} joint consideration of combinations of Histories Theory, 
Records Theory and the Semiclassical Approach (to which I also subscribe \cite{Records, forth}).

\noindent
ii)  Kouletsis' \cite{Kouletsis08} tie between Histories Theory and the Internal Time Approach. 

\noindent
iii) E.g. Dittrich's \cite{Ditt} tie between Rovelli-type Timeless Approaches and the Internal Time 
Approach.  

\noindent
iv) The combination of Type 2 `Rovelli' Timeless Approaches and Histories Theory used by Thiemann (see 
Sec 8.3).

\noindent I have argued that i) [and indeed Histories Theory including ii) and Records Theory] are as 
yet incomplete Problem of Time programs.  

\mbox{ }

\noindent 4) Finally, for the rest of this Section, I consider which of the above Problem of Time 
strategies have counterparts in Affine Geometrodynamics, the Ashtekar Variables approach and 
Supergravity and String/M-theory. 
Such departure from geometrodynamics as the arena for the Problem of Time reflects change in focus in
Quantum Gravity to Ashtekar Variables and String Theory since the mid-80's (or even mid-70's, in the 
case of Supergravity, though much of the earlier work in Supergravity and in String Theory was done 
from rather different perspectives, so that the Problem of Time has in fact has not yet been studied 
that much for these).
Which Problem of Time strategies have been substantially developed for these? 
Do any of these perform substantially differently from in ordinary geometrodynamics?

\subsection{Strategies for the Problem of Time in Affine Geometrodynamics}  

I first note that some Problem of Time strategies are {\it universal} in the sense that they take 
formats that exist no matter what the underlying theory is (thus this point applies in the next two 
subsections as well).   
One case of this is that whatever theory has Semiclassical, \NSI, Conditional Probabilities 
Interpretation, Records, Histories, and `Rovelli' Timeless types of strategies.
Another case of this is that approaches based on mastering the diffeomorphisms are widely blocked from 
progress.
(Though some toy models have simpler entities than the diffeomorphisms, and it may not be clear with 
what the notion of diffeomorphisms is replaced in discrete theories, though these would then be 
required to at least approximately reproduce the notion of diffeomorphism in suitable `spacetime' 
limits.)

In Affine Geometrodynamics \cite{Affine}, the unreduced Hilbert space changes, as does the specific 
form of the WDE, though it remains a frozen equation.  
As regards Tempus Ante Quantum approaches, solving for the time at the classical level is before there 
is any plain--affine distinction.  
There may however still be some scope for different commutation relations subsequently arising. 
As regards Tempus Post Quantum affine approaches, these include an analogue of superspace time (as the 
signature of the wave equation is unaltered by passing to the affine approach).  
As regards the subsequent inner product issue, the potential not respecting the conformal Killing vector 
is an issue that occurs at the classical level, so that the non-useability of a Klein--Gordon inner 
product carries through to the affine case.  
I am not aware of Third Quantization having been tried in this case, but a number of reasons for it  
not appearing to be very promising as a Problem of Time resolution do look to carry over to the 
affine case.  

\noindent
The \NSI should not care about such a change, since this approach does not make use of the WDE.  
As the \CPI does work in a Wheeler--DeWitt framework \cite{K92}, this will change in detail.  
The start of Records Theory involves notions of distance and of information at the classical level, 
which are unaffected by this change, though subsequent changes would then be expected at the quantum 
level.  
A Histories Theory approach to Affine Geometrodynamics was considered by Kessari \cite{Kessari}.
Observables approaches are, in any case, geared toward the Loop Quantum Gravity setting (though it 
would still be interesting to see how these approaches do in the geometrodynamical setting).  
\cite{I93} indicates that affine version of evolving constants of the motion exists.

All in all, one seldom gets far enough \cite{I09-10} in non-formal detail with Problem of Time 
approaches to Geometrodynamics for the ordinary versus affine geometrodynamics distinction to give 
notable changes in behaviour.

\subsection{Strategies for the Problem of Time for GR in Ashtekar Variables} \label{SSSec: AV} 

The classical Hamiltonian constraint now has a rather different form.
\beq
{\cal H} := 
\mbox{tr}(\mbox{\tt E\normalfont}^{i}\mbox{\tt E\normalfont}^{j}\mbox{\tt F\normalfont}_{ij}) = 0 
\label{ashham} \mbox{ } ,
\eeq
with associated linear constraints

\noindent
\beq
{\cal H}_i := \mbox{tr}(\mbox{\tt E\normalfont}^{j} \mbox{\tt F\normalfont}_{ji}) = 0 
\label{ashmom} 
\mbox{ (new form of momentum constraint)} ,
\eeq
\beq
{\cal G}_A := \mbox{\bf\tt D\normalfont}_{i}{\mbox{\tt E\normalfont}^{i}}_A := 
\pa_{i}{\mbox{\tt E\normalfont}^{i}}_A  + [\mbox{\tt E\normalfont}_{i},\mbox{\tt E\normalfont}^{i}]_A 
= 0 \mbox{ (Yang--Mills Gauss constraint) } , \label{ashgauss} 
\eeq
This is like SU(2) Yang--Mills theory (with [\mbox{ },\mbox{ }] here the SU(2) bracket), but with extra 
constraints (\ref{ashham},\ref{ashmom}).  
%
%
The above form is the early complex Ashtekar Variables as opposed to the more modernly used real 
Ashtekar Variables with their Immirzi parameter ambiguity.  
As the Ashtekar Variables Hamiltonian constraint is still of quadratic and not linear form, it too has 
a Frozen Formalism Problem.
Elsewise, this form of the constraint is simpler than the geometrodynamical form of the Hamiltonian 
constraint. 
Some simpleness is lost in passing to real (Barbero) form, but the quadratic and not linear feature 
remains, and thus the Frozen Formalism Problem remains. 
Kucha\v{r} \cite{K93} only provides a critique of the early complex version.  
A Quantum Spin Dynamics version \cite{Thiemann01} and the Master Constraint program \cite{Phoenix, 
Thiemann} (also known as the `Phoenix project') have been developed since.  
The Master Constraint approach seeks in particular to use a Type 2 `Rovelli' Timeless Approach 
as regards dealing with the Problem of Time.
Thus considering such approaches in detail is of importance in determining whether the treatment of the 
Problem of Time that is commonly invoked in the Loop Quantum Gravity context is an adequate one. 
N.B. however that Type 2 `Rovelli' Approaches are not the only Problem of Time strategies pursued in 
Loop Quantum Gravity. 
However, in \cite{Rovellibook} Rovelli is not definitive about such an approach as yet being 
demonstrated to be a fully viable strategy for dealing with the Problem of Time, a point of view that 
Smolin largely appears to agree with \cite{Smolin08}.     
Thus, reliance on Rovelli's book to constitute a resolution of the Problem of Time, as is becoming 
commoner in the literature, is not by itself a currently-justifiable argument.

One should furthermore bear in mind that further work along the lines of the main suggestions in \cite{Rovellibook} 
has also been done by e.g. Thiemann and Dittirch, though it remains unclear, at least to me, whether 
their work makes significant further progress in resolving the (full meaning of the) Problem of Time.  
In particular, Thiemann \cite{Thiemann} vouches for Partial and Complete Observables (of Type 2 
`Rovelli' Approaches being (much of) how the Problem of Time is to be treated in Loop Quantum Gravity.
He gets round at least some of the difficulties faced by approaches of this kind by using his 
{\it Master Constraint Program} \cite{Phoenix}.  
This involves packaging the infinity of constraints into a single constraint known as the {\it master 
constraint},   
\beq
M = \frac{1}{2}\sum_{I, J \in {\cal I}}H_{I}K^{IJ}H_{J}
\eeq
for I, J indexing a bunch of constraints (first class or not) and $K^{IJ}$ a positive operator on the 
space of square-summable sequences over the index set ${\cal I}$.  
Then constraint surface for M coincides with that for $H_I = 0\mbox{ }\forall\mbox{ }I \in {\cal I}$.   
[By 
\beq
\{F,\{F, M\}\}_{M = 0} = \sum_{I, J}\{F,H_I \}_{M = 0}K^{IJ}\{F, H_J\}_{M = 0}
\eeq
the single master equation $\{F,\{F, M\}\}_{M = 0} = 0$ is equivalent to the infinity of equations 
$\{F, H_I\}_{M = 0} = 0 \mbox{ } \forall \mbox{ } I$, so the master equation precisely picks out weak 
%
%
observables.   
This and slightly more work on strong observables sorts out, Thiemann claims, the Problem of Time in 
terms of evolving constants of the motion, which have furthermore a clear physical interpretation in 
terms of Partial Observables.]  
%
%
At the quantum level, however, Thiemann \cite{Thiemann} cautions that what he is doing is to present an 
idea, and one that is strongly tied to the Master Constraint Program for the Loop Quantum Gravity 
approach to quantum GR (which would therefore appear not to be aiming for a universal resolution by 
which Type 2 `Rovelli' Timeless Approaches become fully satisfactory answers to the Problem of Time).  
The Master Constraint Program's considerably superior mathematical machinery does present a much more 
viable quantization approach as compared to the technical problems (7.1-2, 7.1-3) incurred by earlier 
Type 2 `Rovelli' Timeless Approaches.

\mbox{ } 

\noindent Possible Problem 8.3-1. The Master Constraint Program does however have some rather 
non-standard features, so that reformulating simpler toy models in analogous form might be capable of 
dismantling it by showing that they do not lead to the standard results for these \cite{Phoenix}.  

\noindent 
\cite{PhoeTest} is a good start on demonstrating the Master Constraint Program to be in fact OK in 
this respect.\footnote{It 
does not however cover how far one can get with a Master Constraint Program for geometrodynamics.}  

\noindent Possible Problem 8.3-2.  Moreover, it seems rather magical for one to get so much out of 
packaging all of the constraints into a single object via the introduction of some highly arbitrary 
matrix $K$; might one have to pay a price for this through dependence on the choice of $K$ creeping in 
at some later stage of the program?  


\noindent Thiemann \cite{Thiemann} additionally brings in some Histories Theory ideas so as to deal 
with closed system and collapse of the wavefunction issues.  

\noindent So, can the `Phoenix Project' melt away Problem of Time `Ice Dragon'?
Some issues here as as follows. 


\noindent Problem 8.3-3 This approach has Spacetime Reconstruction issues (in particular, 
as regards correctness of classical limit \cite{Thiemann}).

\noindent Problem 8.3-4 The Problem of Observables also remains \cite{Thiemann} (since this is 
primarily a Type 2 `Rovelli' Timeless Approach, this remains as an expected and prominent issue, 
c.f. 7.1-1).

\mbox{ } 
 
Moreover, the Master Constraint Program is only one alternative among various as to what to do about 
the Problem of Time in Ashtekar Variables GR/Loop Quantum Gravity.   
Dittrich's work (\cite{Ditt}, Sec 7.3, \cite{Thie04b}) might here be used otherwise.      
Additionally, Pullin and Gambini have proposed a distinct approach to the Ashtekar Variables 
formulation's Hamiltonian constraint \cite{PGLQG}. 
A phantom matter time was suggested in \cite{Giesel}; however this is again an unphysical type of 
matter (c.f. Problem 2.6-2).
The time usually used in Loop Quantum Cosmology is a matter time.  
But, in parallel with issues raised long ago (see e.g. \cite{IB75}), is this a good 
timefunction, and are the results obtained dependent on making this choice of timefunction?  
[The Multiple Choice Problem could be an issue here.]
Recent work on unimodular gravity \cite{Geiller} serves, among other things, as a check of this 
Multiple Choice issue (earlier work on this in the Ashtekar Variables context is in \cite{Bom}).
To date, we lack a convincing dilational internal time/confomal approach to the Problem of Time in 
Ashtekar Variables (and in any case, Global Problem 2.5-3 still applies in this setting).
More modern semiclassical Loop Quantum Gravity uses coherent states rather than Born--Oppenheimer-type 
Semiclassical Approach \cite{Thiemann03}; nevertheless, Loop Quantum Gravity has had a fair amount of 
difficulty with recovering semiclassicality.   
Loop Quantum Cosmology's \cite{Bojowald} use of semiclassicality as a future boundary condition rather 
than as a property of the real world to be deduced from the theory may also be questionable
Histories Theory in the Histories Projection Operator style has been studied in the case of Ashtekar 
Variables \cite{Savvidou06}.  
Additionally, spin foam approaches could be viewed as path integral preliminaries upon which a 
Histories Theory scheme could be built, though this presently looks to be an open question.

\subsection{Strategies for the Problem of Time in Supergravity}  

How supersymmetry affects the various Problem of Time stategies is an interesting question since 
supersymmetry ameliorates a number of other problems.
That the bracket of two supersymmetry operations is a translation leads to gauged supersymmetric theory 
that contains diffeomorphisms.   
Thus including supersymmetry makes for an interesting extension of the diffeomorphisms and of the 
Dirac Algebra.
E.g. this ought to change the canonical group involved, and the subsequent representation theory.  
On the other hand, this does not remove the difficulties that arise from taking the bracket of two 
Hamiltonian constraints.

In Supergravity, the unreduced Hilbert space changes, as does the specific form of 
the WDE, though it remains a frozen equation.  
I am not aware of any investigations as yet of reference fluid times for Supergravity.
A unimodular approach for Supergravity was proposed in \cite{unim-SUGRA}.  
In fact the idea linked to this cite in SSec \ref{SSec: Facets} was made in the context of Supergravity 
(but, since it appears in fact to be untied to any specific features of Supergravity, I discussed it 
previously in greater generality).  
A new feature here is that the unimodular field is not ad hoc but part of a multiplet. 
Also, the cosmological time is here a dynamical variable determined by fields on a given hypersurface 
rather than depending on the entire history before this hypersurface.  
Another unimodular approach to Supergravity was proposed in \cite{Nish}.

The problems with Superspace Time and with Third Quantization {\sl as a means of solving the Problem 
of Time} are not known to be resolved by passing to Supergravity. 
(\cite{Nicolai99} discusses Third Quantization of a supergravity theory with some parallels to the 
passage from first-quantized String Theory to String Field Theory.) 
The Semiclassical Approach for Supergravity has been considered in \cite{Kiefersugy}.  
I am not aware of any Timeless Approach or Histories Theory work for Supergravity, but this would be 
expected to be doable on grounds of universality (what is not clear without actually doing these is 
whether Supergravity allows for one to get {\sl any further} than GR with such approaches providing 
resolutions of the Problem of Time).

\subsection{The Problem of Time and String/M Theory}

The presence of background structure means that one can take there to be no Problem of Time at the 
level of perturbative String Theory.\footnote{Loop Quantum Gravity 
has a greater degree of background independence than perturbative String Theory. 
Namely, it is independent of background {\sl metric} structure. 
However, genuine philosophical desire to rid oneself of background structure might well require one to 
go as far as having no background topology, too mathematically ambitious for present-day physics as this 
may be.  
(Sum over histories/worldsheets approaches have at least some capacity to incorporate this. 
Thus e.g. Histories Theory and the Third Quantization Approach may have some chance of being able to 
incorporate this feature. 
But there have been some arguments made for considerably more radical approaches in this respect.)}
Moreover, 1) the string effective action gives a GR-like theory and thus possesses a Wheeler-DeWitt 
type equation, at least at scales somewhat larger than the Planck length \cite{Kiefer09}.
\noindent 2) A number of technical as well as conceptual issues then drive one to seek nonperturbative 
background-independent formulations, and then Problem of Time issues resurface. 
See e.g. \cite{Perry} for a recent classical canonical formulation of M-theory.

The membrane form of M-theory certainly has diffeomorphism issues.  
The bosonic string has also been studied as a toy model of geometrodynamics \cite{KoulKu} [this also 
covers the Histories Brackets Approach], being simpler in the sense that its {\it 2-$d$ worldsheet} 
diffeomorphisms are simpler than GR's 4-$d$ spacetime ones.  
It would then be interesting to extend that work to the superstring, the membrane and the supermembrane. 
Another interesting question with Problem of Time implications is what happens to the notion of 
diffeomorphism when one passes from the membrane form of M-theory to the matrix form.

\cite{timeinM} looks to be along the lines of a quantum-level Records Theory, due to its being based on 
finding distances between states in the {\sl quantum} space of states.
The main obtacle here is that the space of states in question is very formidable mathematically (it 
arises as generalization of how $\mathbb{CP}^n$ arises in the study of ordinary mechanics' quantum 
states), so it is unclear how far one can get with this approach.
As such, it may be worth considering how satisfactory this approach is as a Problem of Time approach in 
the given toy model of the space of quantum states in ordinary QT, and then in some progression of 
models of intermediate difficulty. 
(It is also as yet unclear whether the geometrodynamics and Ashtekar Variables cases of such a scheme 
are more or less mathematically involved than these authors' M-theory case.)

The argument about how the existence of Killing vectors poses a global obstruction to hidden times 
looks extendible to the M-theory context.  
Also, reference fluid times would now be inappropriate due to clashing with M-theory's fundamentality; 
the possibility of a unimodular time might however remain open.  
One would not expect there to be much if any profit to applying a semiclassical scheme directly 
to M-theory; it ought to suffice to work with the much more conventional effective string action, or 
for Supergravity with 10 spatial dimensions (itself regarded as a semiclassical limit of M-theory).   
Thus (extensions of) the work mentioned in the preceding SSec should be of interest as an emergent 
approach to the Problem of Time in M-theory (and there could be some scope for Problem of Time results 
for Supergravity carrying over to M-theory.)  
The idea of third-quantizing and/or summing over membrane-histories could well retain their usual merits.  
Some spacetime reconstruction issues in string theory are covered in \cite{Seiberg}.

Furthermore, the new specific forms taken by various approaches to M-theory could well lend themselves 
to suggesting further strategies for resolving the Problem of Time.  
I will briefly comment on two possible such features. 
Firstly, the possibility of theories of this kind having two time dimensions has been considered.  
This would certainly change the nature of most Problem of Time strategies, and indeed of the nature of 
most properties of time, posing challenges for the philosophy of time more generally. 
For the moment I caution that, at the classical level, having two time dimensions has very unfortunate 
effects as regards well-posedness of the field equations, and even as regards the nature of causality.

Secondly, I consider whether the AdS--CFT conjecture may furbish a resolution of the Problem of Time.  
On the one hand, this might be precluded by the CFT boundary amounting to a sufficient vestige of 
background structure (so that this would still be akin to resolving the Problem of Time in an 
asymptotically flat universe by anchoring the notion of time to that notion of time which is 
asymptotically privileged). 
On the other hand, both the mathematics of the correspondence and its physical interpretation may be 
sufficiently different from the situation in asymptotically flat geometrodynamics. 
Then the presence of a privileged time, and unitary evolution, in the CFT would point to 
corresponding notions in the bulk spacetime theory, and it would then be very interesting to work out 
what these are in terms of the language of the bulk spacetime theory itself.  
This would presumably give some sort of internal time, of a form not found by prior investigations 
not making use of the AdS--CFT correspondence, perhaps because of taking a highly nonlocal form 
in the spacetime language, or perhaps because of novelty in the form of the spacetime theory involved.  
However, one might conversely suspect that for sufficiently general bulk spacetime theory, then 
the mismatch between such spacetime theories generically having a Problem of Time and CFT's having 
a privileged notion of time amounts to these theories being of sufficiently different nature that no 
correspondence could bridge between the two. 
Thus, while AdS--CFT might resolve the Problem of Time, the Problem of Time might also render AdS--CFT 
absurd, or at least limit the range of situations for which such a correspondence does exist.  
The most pertinent question then would be whether any models of sufficient generality to 
encompass\footnote{I would take this to include explaining the observations, e.g. the origin of  
microwave background fluctuations or whether there are any observable imprints from higher dimensions.}   
all observational gravitational physics can fit within models that do exhibit an AdS--CFT 
correspondence.

\mbox{ } 

\noindent {\bf Acknowledgments.} 
My wife Claire and my friends Amelia, Sophie and Sophie for support.  
Professors Karel Kucha\v{r}, Chris Isham, Carlo Rovelli and Dr Julian Barbour for discussions. 
Professor Enrique Alvarez for references.  
Professors Belen Gavela, Marc Lachi\`{e}ze-Rey, Malcolm MacCallum, Reza Tavakol and Don Page, 
and Dr Jeremy Butterfield, for help with my career.
Fqxi grant RFP2-08-05 for travel money. 
UAM for funding during the editorial stage of this work.

\mbox{ }

\noindent{\bf\Large Appendix 	A: toy models for the Problem of Time}

\mbox{ } 

\noindent
Full quantum GR is technically difficult, so more progress may be made for toy models \cite{K91, K92, 
I93}.
Toy models vary in which ways they closely resemble GR. 
Such toys have some, but not other of the following crucial features of GR: 
infinite-dimensional and curved configuration spaces, 
indefinite kinetic terms, 
the presence of a linear momentum constraint, 
issues specific to the diffeomorphisms and their canonical split and algebra, 
the presence of a complicated and specific potential term, 
and the high nonlinearity of GR's equations of motion.
Toy models for the Problem of Time should not be linearly ranked due to the many ways in which each  
can possess a partial set of these GR-like features. 
One should beware extrapolation from toy models, due many proposals tractable for a toy model then  
turning out to be specifically tied to non-generalizable features of that toy model. 
Also, the lack of certain GR-like features in some models can render some aspects of the Problem of 
Time trivial, impossible, undemonstrable or absurd therein.  
General rules of thumb concerning pros and cons of toy models for the Problem of Time are as follows. 
\noindent
1) Finite models do not exhibit the Functional Evolution Problem 
\noindent
2) Finite models do not exhibit the worst of the technical QFT problems, e.g. nonrenormalizability.  
Thus, here some of the conceptual issues can be discussed in a technical framework that is 
mathematically well-defined.  
The main limitation of these models is the absence of well-behaved perturbative schemes based around 
them.
\noindent
3) Models in which the dynamical entities do not include some sort of physical space will not have  
diffeomorphism issues or spacetime reconstruction issues.  
\noindent
4) Semiclassical, Timeless and Histories Approaches exist in highly model-independent form, so all 
manner of toy models will possess such schemes.  
\noindent 
5) On the other hand, internal time candidates of some types will be tied to the the structure of the 
particular model in question.  
\noindent 
6) Matter time only really makes sense in gravitational contexts in which the appropriate kind of 
matter can be coupled in (though parts of the calculation do have counterparts in e.g. 
electromagnetic theory).
\noindent 
7) Superspace Time/Klein--Gordon 

\noindent Approach/Third Quantization only make sense for models with indefinite 
kinetic terms.  
\noindent
8) Nonrelativistic particle/ parametrized relativistic field, and models 
that have natural choices of embedding variables have no Multiple Choice Problem.

For toy models consisting of coupled oscillators, stationary states, parametrized nonrelativistic 
particles or fields, and relativistic particles, see \cite{K92} and references therein, since there has 
not been much progress since then from using these.
On the other hand, there have been significant developments concerning toy models that consist of 
2 + 1 gravity, relational particle models (RPM's), bosonic strings, detectors, and information 
gathering and utilizing systems (IGUS) (see below for explanations of what all of these are). 
I discuss these below including minisuperspace and midisuperspace toy models too as these provide 
important comparisons, are certainly still actively used, and have Ashtekar Variables \cite{Bojowald}, 
Supergravity \cite{eath} and stringy scenario counterparts (see e.g. $\cite{GV03}$), though 
midisuperspace modelling of all of these is still in its infancy. 
Ashtekar Variables and String/M-theory have their own additional toy models (it is not clear how much 
these toy models have to say about the Problem of Time).  

\mbox{ }

\noindent{\bf\large A.1 Minisuperspace}\label{SSec: minisuperspace} 

\mbox{ }

\noindent 
This is a well-known toy model of superspace \cite{Mini, eath, Bojowald, GV03}; it involves 
consideration of just the homogeneous spatial 3-geometries.
Thus it is finite, and does not possess a notion of inhomogeneity/structure/localization or a nontrivial 
momentum constraint.
The restriction of the GR supermetric to minisuperspace is a $3 \times 3$ symmetric array rather than 
the full case's $6 \times 6$ one at each space point.
Minisuperspace does inherit specific potentials and kinetic term indefiniteness from GR.

Minisuperspace is used e.g. to illustrate the Klein--Gordon and square-root Schr\"{o}dinger approaches' 
inequivalence\cite{IB75}, simple scale, York and matter internal time models \cite{I93}, and Histories 
Theory approaches \cite{AS05}.  
It is also around as far as most work on Loop Quantum Cosmology gets as regards nontriviality.

\mbox{ } 

\noindent{\bf\large A.2 Strong Gravity}\label{SSec: ToyStrong}

\mbox{ }

\noindent
Strong gravity \cite{strong} can be viewed as the `$G \longrightarrow \infty$' or `$c \longrightarrow 
0$' limit of GR.  
This is the Carrollian limit in which the Lorentzian lightcone becomes squeezed into a line as points 
become entirely isolated due to the zero speed of physical signals (which is the opposite of the 
Galilean limit, in which the Lorentzian lightcone becomes squashed into a plane of simultaneity due to 
the infinite speed of physical signals.)  
Thus strong gravity consists of entirely isolated worldlines, and is most commonly interpreted as 
ultralocal physics (no spatial derivatives).  
Thus ${\cal H}$ loses the Ricci scalar term here.  
$\Lambda$ needs to be retained to ensure that the theory is nontrivial; strictly, one is taking the 
`$G \longrightarrow \infty$' limit while `keeping ${\Lambda}/{G}$ constant' to arrive at theories 
of strong gravity.    
Ultralocal models require a nonstandard form of renormalization \cite{Klauder}, which may be useful 
for gravitational physics and is closely tied to the affine approach to Geometrodynamics.

As regards Problem of Time applications, if one contemplates a Klein--Gordon inner product for strong 
gravity, unlike for GR the potential is of a fixed sign, and thus directly analogous with the standard 
Klein--Gordon equation (or the tachyonic one, depending on the sign of the $\Lambda$ `mass').  
Thus the Klein--Gordon type inner product ought to be normalizable.  
The issue of the positive--negative modes split of states is then resolved because there {\sl is} a timelike superspace 
conformal Killing vector. 
Attempts to expand perturbatively about this by including a small Ricci scalar term have remained  
been mired in difficulties. 
It can also be used as an internal time model.

\mbox{ }

\noindent{\bf\large A.3 Midisuperspace}\label{SSec: ToyMidi}

\mbox{ }

\noindent
More realistic Midisuperspace models (simple infinite models) such as Einstein--Rosen cylindrical 
waves \cite{midicyl, K92}, spherically symmetrical models \cite{midisphe} (including some models with 
shells \cite{Shell} have been studied.  
While these models are infinite-d, inhomogeneous GR models, they possess (usually fairly large) 
isometry groups and so are still rather simpler than the general GR case.

Midisuperspace is very useful for the Problem of Time, albeit at the cost of being the hardest sort of 
toy model to do calculations with.   
E.g. Einstein--Rosen models and spherical models possess extrinsic internal times, and have been used 
for matter time \cite{HKG06, New6} and observables \cite{Ditt} approaches.    

\mbox{ }

\noindent{\bf\large A.4 Perturbations about minisuperspace}\label{SSec: ToyminisuperspacePerts}

\mbox{ }

\noindent
This lies just to the infinite side of the finite--infinite frontier.  
This has been done for linear perturbations (first by \cite{HallHaw} 
though the calculations have been somewhat efficientized since; the corresponding second-order 
calculation remains daunting).

The principal Problem of Time strategy application of this model to date is the Semiclassical Approach 
(see e.g. the critique of \cite{K99}).
It should also be noted that this model suffices at least for some purposes for actual study of the 
origin of galaxies or of microwave background perturbations.  

\mbox{ }

\noindent{\bf\large A.5 2 + 1 gravity}\label{SSSec: Toy2+1}

\mbox{ }

\noindent
See \cite{Carlip} for a large review of quantum gravity in $2+1$ dimensions.   
This is simpler than in 3 + 1 dimensions due to e.g. being flat and having no local degrees of freedom  
(there are still global degrees of freedom).
It does however continue to have a number of nontrivial features of 3 + 1 gravity, in particular it 
retains nontrivial diffeomorphisms.

2 + 1 models are useful e.g. to illustrate nonuniqueness of the resultant quantum thory, internal time, 
breakdown of the saddle approximation underlying the Semiclassical Approach \cite{Carlip}, 
and in testing a number of Loop Quantum Gravity ideas e.g. \cite{QSDIV, PhoeTest}.

\mbox{ }

\noindent{\bf\large A.6 Strings instead of 3-geometries}\label{SSec: ToySrings}

\mbox{ }

\noindent
These are treated in a suitable canonical form in \cite{KuTorre, K92, KoulKu}.    
They are models with a 1-d space of extent. 
Thus, they serve as toy models of some aspects of the diffeomorphisms, 
and also of the History Brackets Approach with a space map as well as a time map. 

\mbox{ }

\noindent{\bf\large A.7 Relational Particle Models (RPM's)}\label{SSec: RPM's}

\mbox{ }

\noindent
RPM's are mechanical theories in which only relative times, relative angles and (ratios of) relative 
separations have physical meaning. 
RPM's have a quadratic energy constraint that manifests the Frozen Formalism Problem.
RPM's also have linear constraints corresponding to rotations [and sometimes scalings]. 
Thus RPM's involve a `relational space' = $\mathbb{R}^{N -1}/\mbox{Rot}(d)$ analogue of 
Superspace($\Sigma$) = Riem($\Sigma$)/Diff($\Sigma$) [and a `shape space' = $\mathbb{R}^{N - 1}/\mbox{Rot}(d) \times \mbox{
Dil}$ analogue of CS($\Sigma$) = Riem($\Sigma$)/Diff($\Sigma$) $\times$ Conf($\Sigma$)].\footnote{$d$ 
is the spatial dimension of the model and $N$ is the total number of particles; it is $N$ -- 1 that 
appears in these configuration spaces due to trivial removal of the translations. Rot($d$) is the group 
of $d$-dimensional rotations, and Dil is the group of dilatations.}
%
RPM's also have a notion of structure/inhomogeneity by which they are a useful toy model of some 
midisuperspace features even if $d$ = 1. 
For $d = 1$ or 2 (for which nontrivial analogy with geometrodynamics still holds), RPM's have simple  
configuration space mathematics: shape space = $\mathbb{S}^{N - 2}$ and $\mathbb{CP}^{N - 2}$ 
respectively, with the corresponding relational spaces being the cones over these (which, in the former 
case, are simply $\mathbb{R}^{N - 1}$).  
This renders these RPM's particularly solvable, making them useful for investigating numerous Problem of 
Time strategies, and comparisons between, and compositions of, these strategies.  
The above spaces become quotiented by $\mathbb{Z}_2$ if one identifies mirror-image shapes; this case 
more closely parallels Affine Geometrodynamics, and, for triangle configurations, amounts to imposing a 
signed area in parallel to the det($h$) $>$ 0 condition.

RPM's with scale included have an Euler time that is a dilational time, i.e. a counterpart of GR's York 
time, and are analogous to Halliwell--Hawking's scheme \cite{HallHaw} for inhomogeneous perturbations 
about homogeneous semiclassical quantum GR.  
Moreover, this now has a rather simpler shape dynamics coupled to the dominant scale dynamics, which is 
useful in qualitative investigations of the appropriateness of the Semiclassical Approach.  
Unlike GR, RPM's are finite, positive-definite, and are highly free in what form their potential takes 
(but this can be largely used up to align the models to share dynamics of scale with Quantum Cosmology).  
RPM's are also a useful arena for investigating Records Theory, due to their possessing readily useable 
notions of locality in space (clumping) and configuration space (from being positive-definite). 
Thus, finally RPM's will be useful for investigation of the combination of Semiclassical, Records and 
Histories Theory ideas (Histories Theory being guaranteed to be formulable for RPM's due to its 
universality).  

\mbox{ }

\noindent{\bf\large A.8 Detector models}\label{Detector}

\mbox{ }

\noindent
Halliwell \cite{H99} coupled such as an up--down detector to a mechanics model, and also separately 
included a harmonic oscillator detector within one's mechanics model.  
This can hold information about one Fourier mode in the signal, thus exemplifying that even very simple 
systems can make imperfect records.   

\mbox{ }

\noindent{\bf\large A.9 Information Gathering and Utilizing Systems (IGUS) or `Model Minds'}\label{IGUS}

\mbox{ }

\noindent
These \cite{IGUS} are possibly a competitor with midisuperspace for maximum realism, from which they 
very greatly differ.  
Does quantum GR admit any solutions that contain such?  
These are a class of toy model that one would like to have for Timeless and Histories Theory Approaches, 
but it is far from clear how to make such models from a practical perspective.



\begin{thebibliography}{99}

\footnotesize 


\bibitem{Wheeler}             J.A. Wheeler, in {\it Battelle Rencontres: 1967 Lectures in Mathematics 
                              and Physics} ed. C. DeWitt and J.A. Wheeler (Benjamin, New York 1968).

\bibitem{K81}                 K.V. Kucha\v{r}, in {\it Quantum Gravity 2: a Second Oxford Symposium} 
                              ed. C.J. Isham, R. Penrose and D.W. Sciama (Clarendon, Oxford 1981).

\bibitem{K91}                 K.V. Kucha\v{r}, in {\it Conceptual Problems of Quantum Gravity}
                              ed. A. Ashtekar and J. Stachel (Birkh\"{a}user, Boston 1991).  

\bibitem{K92}                 K.V. Kucha\v{r}, in {\it Proceedings of the 4th Canadian Conference on 
                              General Relativity and Relativistic Astrophysics}   
                              ed. G. Kunstatter, D. Vincent and J. Williams (World Scientific, Singapore 1992).

\bibitem{I93}                 C.J. Isham, in {\it Integrable Systems, Quantum Groups and Quantum Field 
                              Theories} ed. L.A. Ibort and M.A. Rodr\'{\i}guez (Kluwer, Dordrecht 1993), 
                              gr-qc/9210011.

\bibitem{K99}                 K.V. Kucha\v{r}, in {\it The Arguments of Time} ed. J. Butterfield 
                              (Oxford University Press, Oxford 1999).

\bibitem{Kieferbook}          C. Kiefer, {\it Quantum Gravity} (Clarendon, Oxford 2004).

\bibitem{Rovellibook}         C. Rovelli, {\it Quantum Gravity} (Cambridge University Press, 
                              Cambridge 2004).  

\bibitem{Smolin08}            L. Smolin, Problem of Time Course (2008), available in video form at http://pirsa.org/C08003.

\bibitem{DeWitt}              B.S. DeWitt, Phys. Rev. \bf 160 \normalfont 1113 (1967).

\bibitem{Bubble}              K.V. Kucha\v{r}, J. Math. Phys. {\bf 13} 758 (1972).

\bibitem{UW89}                W. Unruh and R.M. Wald, Phys. Rev. {\bf D40} 2598 (1989). 

\bibitem{Kiefer99}            C. Kiefer, Lect. Notes Phys. {\bf 541} 158 (2000).  

\bibitem{K93}                 K.V. Kucha\v{r} 1993, in {\it General Relativity and Gravitation 1992},  
                              ed. R.J. Gleiser, C.N. Kozamah and O.M. Moreschi M (Institute of 
                              Physics Publishing, Bristol 1993), gr-qc/9304012.

\bibitem{Thiemann}            T. Thiemann, {\it Modern Canonical Quantum General Relativity} (Cambridge 
                              University Press, Cambridge 2007). 

\bibitem{Dirac}               P.A.M. Dirac, {\it Lectures on Quantum Mechanics} (Yeshiva University, New York, 1964).  

\bibitem{I84}                 C.J. Isham, in {\it Relativity, Groups and Topology {II}} 
                              ed. B. DeWitt and R. Stora (North-Holland, Amsterdam 1984).  

\bibitem{TSC}                 E.P. Belasco and H.C. Ohanian, J. Math. Phys. {\bf 10} 1053 (1969);
%
                              R. Bartnik and G. Fodor, Phys. Rev. {\bf D48} 3596 (1993).

\bibitem{Affine}              C.J. Isham \& A. Kakas, Class. Quan. Grav. {\bf 1} 621 (1984); 
%
                              633 (1984).
%
                              J.R. Klauder, Int. J. Geom. Meth. Mod. Phys. {\bf 3} 81 (2006), 
                              arXiv:gr-qc/0507113



\bibitem{Whitrow}               G.J. Whitrow, {\it The Natural Philosophy of Time} (Thomas Nelson, 
                                London 1961).

\bibitem{Reichenbach}           H. Reichenbach, {\it The Direction of Time} (University of California Press, Berkeley and Los Angeles, 1956).

\bibitem{Denbigh}               K.G. Denbigh, {\it Three Concepts of Time} (Spinger-Verlag, Berlin, 1981).


\bibitem{IB75}                W.F. Blyth and C.J. Isham, Phys. Rev. {\bf D11} 768 (1975). 
 
\bibitem{York72}              J.W. York, Phys. Rev. Lett. \bf 28 \normalfont 1082 (1972);
%
                              J. Math. Phys. {\bf 13} 125 (1972);  
%
                              {\bf 14} 456 (1973).

\bibitem{Torre}               C.G. Torre Phys. Rev. {\bf D46}  3231 (1992), hep-th/9204014.  

\bibitem{Kiefer09}            C. Kiefer, Gen. Rel. Grav. {\bf 41} 877 (2009), arXiv:0812.0295. 
  

\bibitem{HK99}                P. H\'{a}j\'{\i}cek and J. Kijowski, Phys. Rev. {\bf D61} 024037 (2000), 
                              gr-qc/99080451.  

\bibitem{T97}                 C.G. Torre, in {\it Gravitation and Cosmology} ed. S. Dhurandhar and 
                              T. Padmanabhan (Kluwer, Dordrecht 1997).

\bibitem{06I}                 E. Anderson, Class. Quantum Grav. {\bf 23} (2006) 2469, gr-qc/0511068. 


\bibitem{Gau}                 K.V. Kucha\v{r} and C.G. Torre,  Phys. Rev. {\bf D43} 419 (1991).

\bibitem{New2}                K.V. Kucha\v{r} and C.G. Torre, Phys. Rev. {\bf D44} 3116 (1991).  

\bibitem{BK94}                J.D. Brown and K.V. Kucha\v{r}, Phys. Rev. {\bf D51} 5600 (1995), 
                              gr-qc/9409001.

\bibitem{K94}                 K.V. Kucha\v{r}, in {\it Directions in General Relativity} ed. B. 
                              Hu, M. Ryan and C. Viveshvara (Cambridge University Press, Cambridge 1993).  

\bibitem{KR95}                K.V. Kucha\v{r} and J.D. Romano, Phys. Rev. {\bf D51} 5579 (1995), 
                              gr-qc/9501005.

\bibitem{BroMa}               J.D. Brown and D. Marolf, Phys. Rev. {\bf D53} 1835 (1996),  
                              arXiv:gr-qc/9509026.  

\bibitem{Markopoulou}         F. Markopoulou, Class. Quantum Grav. {\bf 13} 2577 (1996).

\bibitem{BK97}                J. Bicak and K. Kucha\v{r}, Phys. Rev. {\bf D56} 4878 (1997), 
                              gr-qc/9704053.  

\bibitem{HKG06}               Z. Horvath, Z. Kovacs and L.A. Gergely, Phys. Rev. {\bf D74} 084034 
                              (2006), gr-qc/0605116. 

\bibitem{New6}                Z. Kovacs, L.A. Gergely, Z. Horvath, gr-qc/0612052.  

\bibitem{NewUni}              L. Smolin, arXiv:0904.4841.

\bibitem{UniGryb}             S. Gryb,   arXiv:1003:1973. 



\bibitem{LR79Banks}           V.G. Lapchinski and V.A. Rubakov, Acta Physica Polonica {\bf B10} (1979);  
%
                              T. Banks, Nu. Phys. {\bf B249} 322 (1985).   

\bibitem{HallHaw}             J.J. Halliwell and S.W. Hawking, Phys. Rev. {\bf D31} 1777 (1985).  

\bibitem{Zeh86}               H.D. Zeh, Phys. Lett. {\bf A116} 9 (1986).  

\bibitem{BS}                  J.B. Barbour and L. Smolin, unpublished preprint from 1989.  

\bibitem{Kiefer9394}          C. Kiefer, in {\it FESt-Proceedings on the Concepts of Space and Time} ed. 
                              E. Rudolph and I-O. Stamatescu (Springer-Verlag, Berlin 1993), 
                              gr-qc/9308025;
%
                              in Lect. Notes Phys. {\bf 434} (1994). 
   
\bibitem{Semicl}              A.O. Barvinsky and C. Kiefer, Nu. Phys. {\bf B526} 509 (1998), gr-qc/9711037. 

\bibitem{SemiclI}             E. Anderson, Class. Quantum Grav. {\bf 24} 2935 (2007), gr-qc/0611007.

\bibitem{GVH}                 H. Groenwold, Physica {\bf 12} 405 (1946);
%
                              L. Van Hove, Acad. Roy. Belg. {\bf 37} 610 (1951).



\bibitem{B94I}                J.B. Barbour, Class. Quantum Grav. \bf 11 \normalfont 2853 (1994).

\bibitem{B94II}               J.B. Barbour, Class. Quantum Grav. {\bf 11} 2875 (1994).

\bibitem{EOT}                 J.B. Barbour, {\it The End of Time} (Oxford University Press, Oxford 1999).

\bibitem{Bell}                J.S. Bell, in {\it Quantum Gravity 2.  A Second Oxford Symposium} eds. 
                              C.J. Isham, R. Penrose and D.W. Sciama (Clarendon, Oxford, 1981).  

\bibitem{Mott}                N. Mott, Proc. Roy. Soc. Lon. {\bf A126} 79 (1929).   

\bibitem{HT}                  J.J. Halliwell and J. Thorwart, Phys. Rev. {\bf D65} 104009 (2002), 
                              gr-qc/0201070.    

\bibitem{H03}                 J.J. Halliwell, in {\it The Future of Theoretical Physics and Cosmology} 
                              (Stephen Hawking 60th Birthday Festschrift volume) ed. G.W. Gibbons, 
                              E.P.S. Shellard and S.J. Rankin (Cambridge University Press, Cambridge 
                              2003).  

\bibitem{RWR}                 J.B. Barbour, B.Z. Foster and N. \'{O} Murchadha, Class. Quantum Grav. 
                              {\bf 19} 3217 (2002), gr-qc/0012089; 
%
                              E. Anderson, Stud. Hist. Phil. Mod. Phys. {\bf 38} 15 (2007), 
                              gr-qc/0511069. 

\bibitem{CaLa}                M. Castagnino, Phys. Rev. {\bf D57} (1998) 750, gr-qc/9604034;    
%
                              M. Castagnino and R. Laura, Int. J. Theor. Phys. {\bf 39} (2000) 1737, 
                              gr-qc/0006012.    


\bibitem{H99}                 J.J. Halliwell, Phys. Rev. {\bf D60} 105031 (1999), quant-ph/9902008. 

\bibitem{HalliRec}            J.J. Halliwell, Phys. Rev. {\bf D64} 044008 (2001), gr-qc/0008046.   

\bibitem{JZ}                  E. Joos and H.D. Zeh, Z. Phys. {\bf B59} 223 (1985).

\bibitem{HD}                  J.J. Halliwell and P.J. Dodd, Phys. Rev. {\bf D67} 105018 (2003),  
                              quant-ph/0312068.  

\bibitem{HW}                  J.J. Halliwell and P. Wallden, Phys. Rev. {\bf D73} 024011 (2006), arXiv:gr-qc/0509013.  

\bibitem{ButterBar}           J.N. Butterfield, Brit. J. Phil. Sci. {\bf 53} 289 (2002), gr-qc/0103055.    

\bibitem{Records}             E. Anderson, Int. J. Mod. Phys. {\bf D18} 635 (2009), arXiv:0709.1892;  
%
                              in {\it Proceedings of the Second Conference on Time and 
                              Matter}, ed. M. O'Loughlin, S. Stani\v{c} and D. Veberi\v{c} (University 
                              of Nova Gorica Press, Nova Gorica, Slovenia 2008), arXiv:0711.3174.  

\bibitem{HH}                  J.B. Hartle and S.W. Hawking, Phys. Rev. {\bf D28} 2960 (1983).

\bibitem{Hawking84}           S.W. Hawking, in {\it Relativity, Groups and Topology II} ed. B.S. DeWitt
                              and R. Stora (North-Holland, Amsterdam 1984).

\bibitem{HP8688}              S.W. Hawking and D.N. Page,  Nu. Phys. {\bf B264} 185 (1986);  
%
                              {\bf B298}, 789 (1988).

\bibitem{Page12}              D.N. Page, Int. J. Mod. Phys. D5 583 (1996), quant-ph/9507024; 
%
                              D.N. Page, in {\it Consciousness: New Philosophical Essays} ed. Q. Smith 
                              and A. Jokic (Oxford University Press, Oxford 2002), quant-ph/0108039.

\bibitem{PAOT}                D.N. Page, in {\it Physical Origins of Time Asymmetry} eds. J.J. 
                              Halliwell, J. Perez-Mercader and W.H. Zurek (Cambridge University Press, 
                              Cambridge, 1994).  

\bibitem{PW83}                D.N. Page and W.K. Wootters, Phys. Rev. {\bf D27}, 2885 (1983).                 

\bibitem{Giddings}            S.B. Giddings, D. Marolf and J.B. Hartle, Phys. Rev. {\bf D74} 064018 
                              (2006), hep-th/0512200.  

\bibitem{Pullin}              R. Gambini, R. Porto and J. Pullin, Gen. Rel. Grav. {\bf 39} 1143 
                              (2007), gr-qc/0603090.


\bibitem{GMH}                 M. Gell--Mann and J.B. Hartle, Phys. Rev. {\bf D47} 3345 (1993). 

\bibitem{Hartle}              J.B. Hartle, in {\it Gravitation and Quantizations: 
                              Proceedings of the 1992 Les Houches Summer School} ed. B. Julia and 
                              J. Zinn-Justin (North Holland, Amsterdam 1995),  gr-qc/9304006.

\bibitem{IL}                  C.J. Isham and N. Linden, J. Math. Phys. {\bf 36} 5392 (1995) , gr-qc/9503063.   
 
\bibitem{ILSS}                C.J. Isham, N. Linden, K.N. Savvidou and S. Schreckenberg, J. Math. Phys. 
                              {\bf 39} 1818 (1998), quant-ph/9711031.  

\bibitem{Anastopoulos}        C. Anastopoulos, J. Math. Phys. {\bf 41} 617 (2000), gr-qc/9903026.

\bibitem{AS05}                C. Anastopoulos and K.N. Savvidou, Class. Quantum Grav. {\bf 22} 1841 2005, 
                              gr-qc/0410131. 

\bibitem{SavRest}             K.N. Savvidou, ``Continuous Time in Consistent Histories " (Ph.D. Thesis, 
                              Imperial College London, 1999), gr-qc/9912076;  
%
                              Class. Quant. Grav. {\bf 18} 3611 (2001), gr-qc/0104081;    
%
                              {\bf 21} 615 (2004), gr-qc/0306034;
%
                              {\bf 21} 631 (2004), gr-qc/0306036;
%
                              Braz. J. Phys. {\bf 35} 307 (2005), gr-qc/0412059.

\bibitem{Savvidou1b}          K.N. Savvidou, J. Math. Phys. {\bf 43} 3053 (2002), gr-qc/0104053. 

\bibitem{Marolf}              D. Marolf, Annals Phys. {\bf 236} 374 (1994), hep-th/9308141.  

\bibitem{Kiefer09bis}         C. Kiefer, J. Phys. Conference Series {\bf 174} 012021 (2009).   

\bibitem{KouletsisTh}         I. Kouletsis, ``Classical Histories in Hamiltonian Systems", 
                              (PhD Thesis, University of London, 2000), gr-qc/0108021.    

\bibitem{Kouletsis08}         I. Kouletsis, Phys. Rev. {\bf D78} 064014 (2008)  arXiv:0803.0125.

\bibitem{KoulKu}              I. Kouletsis and K.V. Kucha\v{r}, Phys. Rev. {\bf D65} 25026 (2002), 
                              gr-qc/0108022.

\bibitem{Sorkin03}            See e.g R.D. Sorkin, gr-qc/0309009.  

\bibitem{H09}                 J.J. Halliwell, Phys. Rev. {\bf D80} 124032 (2009), arXiv:0909.2597.   



\bibitem{Rov91a}              C. Rovelli, page 126 in {\it Conceptual Problems of Quantum 
                              Gravity} ed. A. Ashtekar and J. Stachel (Birkh\"{a}user, Boston, 1991).  

\bibitem{Haj91}               P. H\'{a}j\'{\i}\v{c}ek, Phys. Rev. {\bf D44} 1337 (1991).  

\bibitem{Rov91b}              C. Rovelli, Phys. Rev. {\bf D43} 442 (1991).

\bibitem{Rov91c}              C. Rovelli, Phys. Rev. {\bf D44} 1339 (1991).

\bibitem{DeWitt62W86}         B.S. DeWitt, in {\it Gravitation: an Introduction to Current Research} 
                              ed. L. Witten (Wiley, New York 1962);
%
                              W. Wootters, in {\it Fundamental Questions in Quantum Mechanics} ed. 
                              L. Roth and A. Inomata (Gordon and Breach, New York 1986).

\bibitem{Car9091b}            S. Carlip Phys. Rev. {\bf D42} 2647 (1990); 
%
                              Class. Quantum Grav. {\bf 8} 5 (1991).   

\bibitem{BF08}                J.B. Barbour and B.Z. Foster, arXiv:0808.1223.  

\bibitem{RovelliPMP}          C. Rovelli in {\it Physics Meets Philosophy at the Planck scale} ed. C. 
                              Callender and N. Huggett (Cambridge University Press, Cambridge 2001).

\bibitem{0111}                C. Rovelli, gr-qc/0111037; gr-qc/0202079.

\bibitem{Rovelli02}           C. Rovelli, Phys. Rev. {\bf D65} 044017 (2002), arXiv:gr-qc/0110003.

\bibitem{035}                 C. Rovelli, Phys. Rev. {\bf D65} 124013 (2002), arXiv:gr-qc/0110035. 

\bibitem{0503013}             J.M. Pons and D.C. Salisbury, Phys. Rev. {\bf D71} 124012 (2005), 
                              gr-qc/0503013.

\bibitem{AT93}                I.M. Anderson and C.G. Torre, Phys. Rev. Lett. {\bf 70} 3525 (1993), 
                              gr-qc/9302033.  

\bibitem{Hartle96}            J.B. Hartle, Class. Quantum Grav. {\bf 13} 361 (1996), gr-qc/9509037. 

\bibitem{Thiemann01}          T. Thiemann, gr-qc/0110034. 

\bibitem{Thiemann03}          T. Thiemann, Lect. Notes Phys. {\bf 631} 41 (2003), gr-qc/0210094.  

\bibitem{QSDIV}               T. Thiemann, Class. Quant. Grav. {\bf 15} 1249 (1998), arXiv:gr-qc/9705018.

\bibitem{Zeh09}               D. Zeh, arXiv:0908.3780. 

\bibitem{PSS10}               J.M. Pons, D.C. Salisbury and K.A. Sundermeyer, arXiv:1001.2726.  

\bibitem{otherPOObs}          R. De Pietri and C. Rovelli, Class. Quant. Grav. {\bf 12} 1279 (1995), 
                              gr-qc/9406014.  


\bibitem{Klauder}             J.R. Klauder, Commun. Math. Phys. {\bf 18} 307 (1970); 
%
                              Acta Physica Austriaca Suppl. {\bf 8} 227 (1971). 

\bibitem{I09-10}              I thank Professor Chris Isham for discussions on this point.  

\bibitem{Kessari}             S. Kessari, Class. Quant. Grav. {\bf 24} 1303 (2007), gr-qc/0605105.


\bibitem{Ditt}                B. Dittrich, Gen. Rel. Grav. {\bf 38} 1891 (2007), gr-qc/0411013; 
%
                              Class. Quantum Grav. {\bf 23} 6155 (2006), gr-qc/0507106. 

\bibitem{Phoenix}             T. Thiemann,  Class. Quant. Grav. {\bf 23} 2211 (2006), gr-qc/0305080.   

\bibitem{PhoeTest}            B. Dittrich and T. Thiemann,  Class. Quantum Grav. {\bf 23} 
                              1025, gr-qc/0411138;
%
                             1067, gr-qc/0411139;  
%
                               1089, gr-qc/0411140;  
%
                              1121, gr-qc/0411141;  
%
                              1143 (2006), gr-qc/0411142.  

\bibitem{Thie04b}             T. Thiemann, Class. Quantum Grav. {\bf 23} 1163 (2006), gr-qc/0411031.

\bibitem{PGLQG}               R. Gambini and J. Pullin, Class. Quantum Grav. {\bf 17} 4515 (2000), 
                              gr-qc/0008031.

\bibitem{Giesel}              K. Giesel and T. Thiemann, Class. Quantum Grav. {\bf 27} 175009 (2010), 
                              arXiv:0711.0119.

\bibitem{Bom}                 L. Bombelli, W.E. Couch and R.J. Torrence, Phys. Rev. {\bf D44} 2589 (1991).  

\bibitem{Geiller}             D-W. Chiou and M. Geiller, arXiv:1007.0735.

\bibitem{Savvidou06}          K.N. Savvidou, Class. Quant. Grav. {\bf 23} 4133 (2006), gr-gc/0602021.


\bibitem{eath}                P.D. D'Eath, {\it Supersymmetric Quantum Cosmology} (Cambridge University Press, Cambridge 1996).

\bibitem{Kiefersugy}          C. Kiefer, T. Lueck and P. Vargas Moniz, Phys. Rev. {\bf D72} 045006 
                              (2005), gr-qc/0505158.

\bibitem{unim-SUGRA}          R. Graham and H. Luckock, Phys. Rev. {\bf D55} 1943 (1997), gr-qc/9603054.  

\bibitem{Nish}                H. Nishino and S. Raipoot, Phys. Lett. {\bf B528} 259 (2002), hep-th/0107202. 

\bibitem{Nicolai99}           H. Nicolai, J. Astrophys. Astr. {\bf 20} 149 (1999), hep-th/9801090.  

\bibitem{Perry}               D.S. Berman and M.J. Perry, arXiv:1008.1763.

\bibitem{timeinM}             V. Jejjala, M. Kavic and D. Minic
                              Int. J. Mod. Phys. {\bf A22} 3317 (2007), arXiv:0706.2252.  

\bibitem{Seiberg} N. Seiberg, in {\it Brussels 2005, The Quantum Structure of Space and Time} 163 
(Princeton, Inst. Advanced Study 2006), hep-th/0601234.



\bibitem{GV03}                M. Gasperini and G. Veneziano, Phys. Rept. {\bf 373} 1 (2003), 
                              arXiv:hep-th/0207130.   

\bibitem{Mini}                C.W. Misner, Phys. Rev {\bf 186} 1319 (1969);
                              in {\it Relativity} (Proceedings of the Relativity Conference in the 
                              Midwest, held at Cincinnati, Ohio June 2-6, 1969) 
%
                              ed. M. Carmeli, S.I. Fickler and L. Witten (Plenum, New York 1970); 
%
                              in {\it Magic Without Magic: John Archibald Wheeler} 
                              ed. J. Klauder (Freeman, San Fransisco 1972);
%
                              M. Ryan, {\sl Hamiltonian Cosmology} Lec. Notes Phys. {\bf 13} 
                              (Springer, Berlin 1972); 
%
                              J.B. Hartle and S.W. Hawking, Phys. Rev. {\bf D28} 2960 (1983);
%
                              D.L. Wiltshire, in {\it Cosmology: the Physics of the Universe} ed. B. 
                              Robson, N. Visvanathan and W.S. Woolcock (World Scientific, Singapore 
                              1996) p 473, gr-qc/0101003.  

\bibitem{Bojowald}            M. Bojowald, Living Rev. Rel. {\bf 8} 11 (2005), gr-qc/0601085.  

\bibitem{strong}              C.J. Isham, Proc. R. Soc. Lond. A. \bf 351 \normalfont 209 (1976);
%
                              C. Teitelboim, Phys. Rev. {\bf D25} 3159 (1982); 
%
                              E. Anderson, Gen. Rel. Grav. {\bf 36} 255 (2004), gr-qc/0205118; 
%
                              T. Damour, M. Henneaux and H. Nicolai, 
                              Class. Quant. Grav. {\bf 20} R145 (2003), hep-th/0212256.  

\bibitem{KuTorre}              K.V. Kucha\v{r} and C. Torre, in {\it Conceptual Problems of Quantum 
                               Gravity}, ed. A. Ashtekar and J. Stachel (Birkh{\"a}user, Boston 1991), 
                               p. 326.

\bibitem{Carlip}              S. Carlip, {\it Quantum Gravity in 2 + 1 Dimensions} 
                              (Cambridge University Press, Cambridge 1998).  

\bibitem{BB82}                J.B. Barbour and B. Bertotti, Proc. Roy. Soc. Lond. {\bf  A382} 295 (1982). 

\bibitem{B03}                 J.B. Barbour, Class. Quantum Grav. \textbf{20} 1543 (2003), gr-qc/0211021.

\bibitem{RPMCONCAT}           E. Anderson, Class. Quantum Grav. {\bf 25} 025003 (2008), arXiv:0706.3934;
%
                              {\bf 26} 135021 (2009) gr-qc/0809.3523;
%
                              E. Anderson and A. Franzen, Class. Quantum Grav. {\bf 27} 045009 (2010), 
                              arXiv:0909.2436;
%
                              E. Anderson, arXiv:0909.2439; 
%
                              arXiv:1003.4034;
%
                              arXiv:1005.2507.

\bibitem{SemiclIII}          E. Anderson, arXiv:1101.4916.  

\bibitem{forth}               E. Anderson, forthcoming.

\bibitem{midicyl}             K.V. Kucha\v{r}, Phys. Rev. {\bf D4}, 955 (1971). 
  
\bibitem{midisphe}            K.V. Kuchar, Phys. Rev. {\bf D50} 3961 (1994), arXiv:gr-qc/9403003. 

\bibitem{Shell}               P. H\'{a}j\'{\i}\v{c}ek, Lect. Notes Phys. {\bf 631} 255 (2003), 
                              gr-qc/0204049. 

\bibitem{IGUS}                J.B. Hartle, Am. J. Phys. {\bf 73} 101 (2005), gr-qc/0403001.

\bibitem{GRRM}                George R.R. Martin, {\it Dreamsongs} (Orion, London 2003).  

\end{thebibliography}
\end{document}